\documentstyle[12pt]{article}
\newcommand{\beq}{\begin{equation}}
\newcommand{\eeq}{\end{equation}}

\def\half{{\textstyle{1\over2}}}
\def\quart{{\textstyle{1\over4}}}

\def\kap{{\textstyle{1\over{\kappa^2}}}}
\def\gap{{\textstyle{1\over{g^2}}}}
\def\half{{\textstyle{1\over2}}}
\def\third{{\textstyle{1\over3}}}
\def\thirdf{{\textstyle{{{1}\over{3!}}}}}
\def\quart{{\textstyle{1\over4}}}
\def\eigth{{\textstyle{{{1}\over{8}}}}}
\def\eigthf{{\textstyle{{{1}\over{8!}}}}}

\def\sixthf{{\textstyle{{{1}\over{6!}}}}}
\def\qfac{{\textstyle{{{1}\over{q!}}}}}
\def\twothird{{\textstyle{2\over3}}}
\def\threehalf{{\textstyle{3\over2}}}

\def\tquart{{\textstyle{3\over4}}}

\def\p1half{{\textstyle{{{p+1}\over{2}}}}}

\def\23phalf{{\textstyle{{{23-p}\over{2}}}}}

    \let\p=\pi

 \def\bd{\begin{document}} \def\ed{\end{document}}
\def\ds{\documentstyle} \let\fr=\frac \let\bl=\bigl \let\br=\bigr
\let\Br=\Bigr \let\Bl=\Bigl
\let\bm=\bibitem
\let\na=\nabla
\let\pa=\partial \let\ov=\overline
\newcommand{\be}{\begin{equation}}
\newcommand{\ee}{\end{equation}}
\def\ba{\begin{array}}
\def\ea{\end{array}}
\def\ft#1#2{{\textstyle{{\scriptstyle #1}\over {\scriptstyle #2}}}}
\def\fft#1#2{{#1 \over #2}}
\def\del{\partial}
\def\sst#1{{\scriptscriptstyle #1}}
\def\oneone{\rlap 1\mkern4mu{\rm l}}
\def\ie{{\it i.e.\ }}
\begin{document}
\thispagestyle{empty}
\begin{titlepage}

\bigskip
\hskip 3.7in{\vbox{\baselineskip12pt
}}

\bigskip\bigskip\bigskip\bigskip
\centerline{\large\bf Hidden Symmetry Unmasked:}
\centerline{\large\bf Matrix Theory and $E_{11}$$=$$E_8^{(3)}$}

\bigskip\bigskip
\bigskip\bigskip
\centerline{\bf Shyamoli Chaudhuri
\footnote{E-mail: shyamoli@thphysed.org}}

\medskip
\medskip
\medskip

\centerline{214 North Allegheny Street}
\centerline{Bellefonte, PA 16823, USA}

\bigskip

\date{\today}

\bigskip\bigskip
\begin{abstract}

\noindent Dimensional reduction of eleven-dimensional supergravity
to zero spacetime dimensions is expected to give a theory
characterized by the hidden symmetry algebra $E_{11}$, the
end-point of the Cremmer-Julia prediction for the sequence of
dimensional reductions of 11d supergravity to spacetime
dimensions. In recent work, we have given a prescription for the
spacetime reduction of a supergravity-Yang-Mills Lagrangian with
large $N$ flavor symmetry such that the local symmetries of the
continuum Lagrangian are preserved in the resulting reduced matrix
Lagrangian. This new class of reduced matrix models are the basis
for a nonperturbative proposal for M theory we have described in
hep-th/0408057. The matrix models are also characterized by hidden
symmetry algebras in precise analogy with the Cremmer-Julia
framework. The rank eleven algebra $E_{11}$$=$$E_8^{(3)}$ is also
known as the very-extension of the finite-dimensional Lie algebra
$E_8$. In an independent stream of work (hep-th/0402140), Peter
West has provided evidence which supports the conjecture that M
theory has the symmetry algebra $E_{11}$, showing that it
successfully incorporates both the 11d supergravity limit, as well
as the 10d type IIA and type IIB supergravities, and inclusive of
the full spectrum of Neveu-Schwarz and Dirichlet pbranes. In this
topical review, we give a pedagogical account of these recent
developments also providing an assessment of the insights that
might be gained from linking the algebraic and reduced matrix
model perspectives in the search for M theory. Necessary
mathematical details are covered starting from the basics in the
appendices.
\end{abstract}
\end{titlepage}

\section{Introduction}

In recent works \cite{mat1,mtheory}, we have presented a
nonperturbative proposal for M theory based on a new class of
reduced supermatrix models. We give a new prescription for the
{\em spacetime reduction} of a supergravity theory with large $N$
flavor symmetry and in generic curved spacetime background, which
preserves the local symmetries of the continuum field theory in
the resulting matrix Lagrangian. The models are characterized by
extended symmetry algebras reminiscent of the hidden symmetry
algebras of dimensionally reduced supergravity theories. We also
address in our framework the converse phenomenon, namely, the
emergence of a continuum spacetime in the large $N$ limit of the
reduced matrix model \cite{mtheory}. In an independent, and very
interesting, stream of recent developments
\cite{west1,west2,w0,w1,w2,w4,w5,kw}, West and collaborators have
given convincing evidence that the global symmetry algebra of the
supergravities with 32 supercharges is $E_{11}$\footnote{Appendix
A contains a pedagogical introduction to the infinite-dimensional
affine-, over-, and very-extensions of a finite-dimensional Lie
algebra. The over and very extensions have a Cartan matrix with
Lorentzian signature. Appendix B is a synopsis of the key evidence
for $E_{11}$ as the symmetry algebra of theories with 32
supercharges as given in \cite{w4}.}, a result which holds for
both 11d supergravity {\em and} the 10d type IIA and IIB
supergravities. $E_{11}$ is further conjectured to be the symmetry
algebra of M theory \cite{w2,w4}. In this paper, we give a
pedagogical review and assessment of some of these developments,
stressing the insights into reduced matrix models that might be
gained by exploration of the algebraic perspective.

\vskip 0.1in It is well-known that the toroidally-compactified
eleven-dimensional supergravity, as well as the ten-dimensional
type I-I$^{\prime}$, type IIA-IIB, and the heterotic
$E_8$$\times$$E_8$ and $Spin(32)/{\rm Z}_2$, string
supergravities, exhibit extended global symmetries as a
consequence of the presence of massless scalar fields in the
dimensionally-reduced supergravity
Lagrangian.\footnote{Dimensional reduction is sometimes
distinguished from toroidal compactification by the neglect of the
Kaluza-Klein modes. As is conventional in the string literature,
we will include all of the massless scalars when identifying the
relevant global symmetry group of the theory, irrespective of
their origin.} In 1978, Cremmer and Julia noticed that the
dimensional reduction of a $(D$$+$$n)$-dimensional theory
containing gravity to $D$ dimensions necessarily results in the
appearance of an $SL(n,{\bf R})$ global symmetry, as viewed from
the perspective of the $D$-dimensional spacetime \cite{cj}. This
symmetry is manifest in the form of the dimensionally-reduced
Lagrangian. Including an overall scaling of the volume of the
compactification manifold, the global symmetry group of the
Lagrangian takes the precise form $GL(n, {\bf R})$$\sim$$SL(n,
{\bf R})$$ \times $${\bf R}$; the ${\bf R}$ factor is, therefore,
a {\em hidden} symmetry of the dimensionally-reduced
Lagrangian.\footnote{I should perhaps remind the reader at the
outset that the $n$-dimensional Lorentz algebra is contained
within $GL(n,{\bf R})$, but not within its $SL(n,{\bf R})$
subalgebra. However, it is the volume preserving $SL(n,{\bf R})$
that is relevant to the spacetime reduction of a field theory to a
single spacetime point: it is only in the large $N$ limit of the
resulting zero-dimensional matrix model, that both a continuum
spacetime, and the full Lorentz algebra, are expected to become
manifest. I thank Andrei Mikhailov for requesting this
clarification.} Recall that eleven-dimensional supergravity is one
of the field theoretic low energy limits of M theory. In
\cite{cj}, Cremmer and Julia conjectured, with partial proof, that
the dimensional reduction of 11d supergravity to a Lagrangian in
$11$$-$$n$ dimensions would result in the appearance of the hidden
symmetry group $E_{n}$. For $n$ $\ge$ $3$, this conjecture has
since been verified by direct field-theoretic duality
transformations on the fields in the dimensionally-reduced
classical supergravity Lagrangian \cite{cjlp}.

\vskip 0.1in When the $(D$$+$$n)$-dimensional gravity theory also
contains antisymmetric tensor field strengths, dimensional
reduction will give rise to axionic scalar fields in $D$
dimensions. A global ${\bf R}$ symmetry in a
toroidally-compactified supergravity corresponds to a shift
symmetry of an axion \cite{cj,cjlp}. As an example, consider the
case of eleven-dimensional supergravity with $32$ supercharges.
Upon dimensional reduction, the metric contributes $(11-n)$
dilatonic scalars arising from its diagonal component, and $\half
(11-n)(10-n)$ axionic scalars, ${\cal A}^i_{[0]j}$. The three-form
gauge potential contributes gauge potentials, $({\cal A}_{[3]},
{\cal A}_{[2]i}, {\cal A}_{[1]ij})$, in $D$ dimensions, in
addition to $q$$=$${{1}\over{6}} (11-n)(10-n)(9-n)$ axionic
scalars, ${\cal A}_{[0]ijk}$. Associated with these axions are $q$
shift symmetries, enhancing the global symmetry group to
$GL(n,{\bf R}) \times {\bf R}^q$, where $q$ $=$
$\{0,0,0,1,4,10,20,35,56\}$ in $D$ $=$ $\{ 11,10,9,8,7,6,5,4,3\}$.
The maximal ${\bf R}$ symmetry is realized by all of the new
axions in $D$ dimensions that did not exist in $D+1$ dimensions
\cite{fre}. We emphasize that this conclusion holds {\em prior} to
performing any dualizations: if we were to dualize all axions to
antisymmetric tensor gauge potentials, the Lagrangian would only
exhibit a $GL(n,{\bf R})$ symmetry. In general, there is
considerable freedom to alter the precise enlargement of the
global symmetry group by invoking appropriate field-dualizations
\cite{cjlp}. But we can safely conclude that the volume-preserving
factor, $SL(n,{\bf R})$, will always be a subgroup of the global
symmetry group of the dimensionally-reduced gravity theory. Unlike
the case of rigid large $N$ Yang-Mills theories, however, the
straightforward dimensional reduction of a locally symmetric
theory to $D$ $\le$ $2$ dimensions can be fraught with ambiguity
in distinguishing scalars and gauge potentials. Thus, a more
algebraic perspective on the process of spacetime reduction seems
called for \cite{julia,nic87,cl,ns,mat1,west1,ganor}.

\vskip 0.1in Guided by the observation that there is considerable
freedom to alter the hidden symmetry group of a supergravity
theory by appropriate field-dualizations \cite{cjlp}, it is of
interest to ask what precise enhancement of $SL(10, {\bf
R})$$\times$$G$, where $G$ is the finite-dimensional Yang-Mills
group, determines the symmetry group of Matrix Theory? In this
paper, we investigate this issue by consolidating insights from
the recent works of many authors on the subject of the hidden
symmetries of M theory, and of other supergravity theories
\cite{cjlp,lp,nicolai,west2,w4,eh,kw,ganor,nk}.

\vskip 0.1in If we continue Cremmer and Julia's sequence of
dimensional reductions of 11d supergravity to lower dimensions ro
its logical endpoint, namely, to {\em zero} spacetime dimensions,
we have the prediction $E_{11}$ for the hidden symmetry group. The
notion of spacetime reduction introduced by us in \cite{mat1}
addresses the symmetries of the Lagrangian obtained in this
extreme limit: we studied the dimensional reduction of a
higher-dimensional supergravity theory to {\em zero} spacetime
dimensions, and also to a single spacetime point. The latter
feature was precisely as in Eguchi and Kawai's prescription for
planar reduction \cite{ek}. The reason our prescription for
spacetime reduction recovers a nontrivial large $N$ matrix model,
even in the {\em absence} of Yang-Mills gauge fields in the
higher-dimensional continuum field theory, is that our starting
point is a supergravity Lagrangian with an additional large $N$
{\em flavor} group \cite{mtheory}.

\vskip 0.1in Planar reduction was first applied to the bosonic
rigid large $N$ Yang-Mills theory by Eguchi and Kawai in 1980
\cite{ek}. Dimensional reduction of a rigid $U(N)$ Yang-Mills
gauge theory to a single spacetime point gives what is known as a
reduced unitary matrix model: naively, we set to zero all
spacetime derivatives in the Yang-Mills action, retaining the
$U(N)$ trace of the square of the commutator of $N$$\times$$N$
unitary matrices. The Lagrangian gives a zero-dimensional unitary
matrix model with a quartic self-interaction.\footnote{We use the
term {\em Lagrangian} for a reduced matrix model as follows: the
Feynman path integral describing the quantum mechanics of the
matrix model is a sum over matrix configurations, weighted by an
exponentiated matrix-valued function. In analogy with field
theory, we will refer to the exponentiated function defined on the
field of unitary matrices, as the matrix {\em Lagrangian}. By
contrast, in the case of the famous c=1 matrix model, the matrix
variables were taken to be functions of an auxiliary parameter
later identified with target-space time \cite{mm}. Thus, the
exponentiated function weighting the Feynman path integral in the
case of the c=1 matrix model is the {\em action}, expressed as a
one-dimensional integral over time.} Reduced matrix models arise,
therefore, as the result of a dramatic thinning of the infinite
number of degrees of freedom of a quantum field theory upon
dimensional reduction of all spacetime fields to a single
spacetime point. Remarkably, planar reduced matrix models are
found to share many features of exactly solvable unitary matrix
models. It should be emphasized that many of the notions familiar
from continuum quantum field theory, such as renormalization,
universality classes, vacuum structure, and spontaneous symmetry
breaking, have their counterpart in the matrix models that follow
from spacetime reduction. Likewise, {\em super}matrix models are
obtained when one dimensionally reduces a rigid supersymmetric
large $N$ Yang-Mills theory to a spacetime point. Such supermatrix
models have been the basis of previous conjectures for
nonperturbative string/M theory \cite{bfss,ikkt}.

\vskip 0.1in With the discovery of Dirichlet-pbranes by Dai, Leigh
and Polchinski \cite{dlp}, and with their crucial role as
solitonic carriers of dual electric-magnetic charge in the type I
and type II string supergravities clarified by Polchinski in
\cite{dbrane}, the dimensional reduction of rigid Yang-Mills
theories has found an alternative, and rather interesting, new
interpretation. Recall that in open and closed string theories,
$n$ successive T-duality transformations on $n$ spacetime
coordinates parallel to the worldvolume of $N$ coincident D9branes
in the type IB string theory carrying 10d nonabelian Yang-Mills
gauge fields: $R_n$ $\to$ $\alpha^{\prime}/R_n$, where $n$ $\le$
$10$, converts $n$ components of the worldvolume gauge bosons to
the $n$ components of a scalar field in the $n$-dimensional
spatial bulk orthogonal to the D(9-n)brane \cite{dlp}. The vacuum
expectation values of the $n$ components of the scalar field can
be interpreted as the coordinate locations of the D(9-n)brane
soliton in an $n$-dimensional space. In open string theory, this
scalar excitation has as vertex operator $(\partial_{z}
X^{i}_{\mu} )^2$, where $\mu$$=$$1$, $\cdots$, $n$, and $i$$=$$1$,
$\cdots$, $N$. As first noted by Witten, this implies the
tantalizing fact that the \lq\lq coordinates" of space orthogonal
to the $N$ D(9-n)branes arise as the $N$ eigenvalues of $n$
noncommuting, $N$$\times$$N$, unitary matrices. For example, with
$9$ spatial dualizations, we have $9$ collective coordinates for
the $N$ coincident D0brane solitons: $A^{i}_{\mu}(x^0)$
$\leftrightarrow$ $X^{i}_{\mu} (x^0)$, $i$ $=$ $1$, $\cdots$, $N$,
and $\mu$ $=$ $1$, $\cdots$, $9$, where $x^0$ is time. Here, $i$
is the Chan-Paton index, and the gauge group realized on $N$
coincident D0branes is the nonabelian group $U(N)$, of rank $N$.
In the unoriented type I string theory we obtain, instead, the
orthogonal group $SO(2N)$ as worldvolume gauge group
\cite{polbook}.

\vskip 0.1in More generally, the $X^{\mu}$ coordinate location of
the $i$th D0brane is the $i$th eigenvalue of the $U(N)$ matrix
$X^i_{\mu}$, $i$$=$$1$, $\cdots$, $N$, described above.
Restricting to the Yang-Mills field theory on the one-dimensional
worldvolume of the D0branes, we have a worldvolume Lagrangian that
agrees precisely with the dimensional reduction of the 10d
nonabelian Yang-Mills Lagrangian. This gives the familiar quartic
interaction for one-dimensional $N$$\times$$N$ matrices
\cite{dlp,witnc}. Such matrix Hamiltonians describe the quantum
mechanics of, time-dependent, large $N$ unitary matrices, as in
the Banks-Fischler-Shenker-Susskind proposal for M(atrix) Theory
\cite{bfss}. Planar reduced matrix models, akin to the
Ishibashi-Kawai-Kitazawa-Tsuchiya IIB Matrix Model \cite{ikkt},
follow as the result of taking this logic one step further: we
must T-dualize all {\em ten} directions of spacetime. The
coordinates $X^i_{\mu}$, with $i$$=$$1$, $\cdots$, $N$, and $\mu$
$=$ $0$, $\cdots$, $9$, can now be interpreted as the locations of
$N$ {\em Dinstanton events} in a bulk ten-dimensional spacetime.
Recall that the tension of a Dinstanton has mass dimension zero.
Thus, such a matrix Lagrangian has no dimensionful couplings and
is reminiscent of a topological theory.

\vskip 0.1in We should emphasize that the spacetime interpretation
of reduced unitary matrix model Lagrangians we have just reviewed
views the D(9-n)branes as semi-classical solitons in an embedding
$n$-dimensional spacetime. But what dynamical mechanism is
responsible for generating the spacetime manifold itself? To
address this puzzle, we must delve further into the search for a
matrix formulation of M theory. We require a nonperturbative
formalism for a fundamental theory of the Universe which addresses
the origin of both the long distance interactions {\em in} an
embedding spacetime geometry, as well as the generation of the
background geometry itself. Such a theory would capture the full
spirit of Einstein gravity: matter and spacetime geometry are set
on an equal footing.

\vskip 0.1in It should be noted that both M(atrix) Theory
\cite{bfss}, and the IIB Matrix Model \cite{ikkt}, are conjectured
theories of {\em induced} gravity: linearized gravity appears as
an effective long-distance interaction of the fundamental,
pointlike, degrees of freedom, respectively, D0branes or
Dinstantons, living in an embedding flat spacetime background.
Reconstructing the full nonlinear structure of Einstein gravity
from this simplified starting point has proven prohibitively
difficult \cite{plefka}, as has the problem of extending the
matrix model formalism to curved spacetime geometries
\cite{taylor}. As emphasized by Nicolai \cite{nicolai}, there is
also no evidence in either matrix model conjecture of the
well-established global symmetries of the Einstein supergravities.
It was natural to suspect that the dimensional reductions of {\em
locally} supersymmetric Yang Mills theories would be a more
relevant direction to explore in the context of conjectures for M
theory. Although a concrete suggestion to this effect was made by
Nicolai in 1997-98 \cite{nicolai}, it appears not to have
attracted much attention in the subsequent research literature.
Notice, however, that it is not immediately obvious why performing
the planar reduction of a locally supersymmetric gauged large $N$
Yang-Mills theory, as in \cite{ek}, should give a matrix model
with nontrivial new large $N$ dynamics. In fact, it will become
necessary to modify the Eguchi-Kawai prescription of planar
reduction \cite{ek} in order to preserve the local symmetries of a
gravitational theory with spinors, in generic curved spacetime
backgound, in a corresponding reduced matrix model.

\vskip 0.1in In \cite{mat1,mtheory}, we applied the simple
procedure of planar reduction to a supergravity Lagrangian with
large $N$ flavor group, and both with, and without, a Yang-Mills
gauge sector. We presented our analysis using as prototype the
manifestly supersymmetric 10d Lagrangian density obtained in the
low energy limit of the heterotic string theory, computed up to
quartic order in the $\alpha^{\prime}$ expansion in \cite{br2,br},
and inclusive of gauge-coupling dependent corrections required by
closure of the supersymmetry algebra. The resulting planar reduced
matrix models with previously studied matrix models, as summarized
in Appendix C of this paper. Our discussion includes an especially
elegant and simple result for the planar reduction of the 11d
supergravity Lagrangian. We explain why simple planar reduction
always results in the absence of any remnant of the spectrum of
supergravity pform potentials in the corresponding reduced matrix
model, despite our introduction of a large $N$ flavor symmetry in
order to obtain a nontrivial matrix model.

\vskip 0.1in Appendix C also contains a summary of our modified
prescription for {\em spacetime reduction}, explaining how the
local symmetries of the continuum Lagrangian can thereby be
preserved in the reduced matrix model. The key insight is to
recognize that the Lagrangian density in quantum field theory
satisfies locality: thus, the spacetime reduction of all spacetime
fields to {\em linear} forms defined on the infinitesimal patch of
local tangent space at a single spacetime point, suffices to
preserve all of the local symmetries of the continuum Lagrangian
in a corresponding reduced matrix model. We then explain the
mechanism for {\em spacetime emergence} as the eigenvalue
coordinates of the zehn(elf)bein matrix array in the large $N$
limit, demonstrating self-consistency with the basic relations of
Riemannian geometry. We exhibit the form of infinitesimal
supersymmetry transformations, and of field redefinitions, under
spacetime reduction. Notice that the large $N$ flavor symmetry has
been chosen to commute with the local symmetries of the
Lagrangian, namely, Lorentz, supersymmetry, and gauged Yang-Mills
transformations.

\vskip 0.1in In section 3, we review the description of the global
symmetry algebra of the nonmaximal 10d supergravity theory with a
nontrivial Yang-Mills sector \cite{w5}, placing it within the
larger context of theories with sixteen supercharges
\cite{chl,cp,mat1,mtheory}. The full details of the precise
continuum Lagrangian of interest to us, namely, that of the
circle-compactified type I-I$^{\prime}$-massive IIA-IIB-heterotic
theory, where all six different string theory limits of this
theory with sixteen supercharges are obtained by suitable field
redefinitions and target-duality transformations alone
\cite{berg}, are as yet unknown. But it is evident that the
methodology for the derivation of the relevant reduced supermatrix
Lagrangian is clear. The {\em bosonic} sector of the supergravity
Lagrangian of interest to us has been well-studied in the
literature, including our previous works \cite{flux,cosmo}. In
particular, the full spectrum of supergravity pform potentials has
been shown to appear within the Lagrangian framework \cite{berg}.
Notice that electric, and magnetic dual Dp(6-p)brane pairs, with
$-2$$\le$$p$$\le$$9$, are represented on an equal footing, as in
the worldsheet formalism of perturbative string theory
\cite{cosmo}, and as corroborated by the analysis of the global
symmetry algebra of the massive IIA supergravity given by
Schnackenburg and West \cite{w2}, reviewed in section 2 of this
paper. Since the self-dual nature of this theory is bound to
introduce some subtleties in the form of the full supersymmetric
Lagrangian \cite{berg,cjlp2}, including all of the fermionic terms
required by closure of the supersymmetry algebra, as in
\cite{br2,br}, we do not present a definitive matrix model
Lagrangian in this section. Rather, we discuss the important issue
of incorporating generic backgrounds, reviewing some established,
but less widely-known, facts about the nature of the vacuum
landscape of theories with sixteen supercharges. This is the
broad-brush picture that has emerged from the detailed study of
CHL models \cite{chl,cp}, and more generic classes of flux
compactifications \cite{ps,cl,others} in recent years.

\vskip 0.1in How does our proposal relate to recent studies of the
hidden symmetry algebra of M theory? We address this question in
Section 4, emphasizing how the algebraic framework can lend
significant insight into some key aspects of our proposal. In
particular, we present a conjecture for the emergence of theories
with 32 supercharges, and no Yang-Mills sector, as a special limit
of the theory with sixteen supercharges in the algebraic
framework. Concrete evidence for self-duality in the worldsheet
formalism of perturbative string theory has been given by us in
\cite{path,flux,cosmo}, a work done in partial collaboration with
Chen and Novak, building on the earlier results in \cite{cmnp}. We
present the worldsheet computation of the tension of a D(-2)braneb
coupling to a (-1)form supergravity potential, the magnetic dual
of the nine-form potential of massive IIA supergravity. In this
paper, we note the corroborating evidence for self-duality
presented by Schnakenburg and West in their analysis of the global
symmetry algebra of the massive IIA supergravity \cite{w1}. We
review West's arguments in favor of the very-extended Lorentzian
Kac-Moody algebra $E_{11}$ as the hidden symmetry algebra of the
ten and eleven dimensional supergravities with 32 supercharges in
Appendix B \cite{w4}. We conclude with a list of open questions,
including those presented in \cite{mtheory}, and outline some key
directions for future work.

\section{Dualizations, Self-duality, and the (-1)form Potential}

Let us summarize some of the key insights gained in recent studies
of hidden symmetry groups in supergravity. In the original work
\cite{cj}, Cremmer and Julia pointed out that, upon a Weyl
rescaling to the Einstein frame metric, the ${\bf R}$ subgroup of
$GL(n,{\bf R})$ becomes a {\em hidden symmetry}: it is no longer
manifest in the Einstein frame Lagrangian. Based on the counting
of massless scalar fields in succeeding dimensions, it was
conjectured that the hidden symmetry group of the reduced
supergravity in $11-n$ dimensions would take the general form
$E_{11-n}$. The details were worked out for the dimensional
reduction to four dimensions, establishing the appearance of the
left coset scalar manifold $E_7$/$SU(8)$. $SU(8)$ is the maximal
compact subgroup of $E_7$ of identical rank. Cremmer and Julia
pointed out that the appearance of a coset structure $G/H$, where
$G$ is a noncompact internal symmetry group and H is the compact
local invariance group, was a generic consequence of dimensional
reduction, implying that the Lagrangian for supergravity scalars
always takes the form of a nonlinear realization of a finite
semi-simple Lie algebra G \cite{ccwz}. The compact local
invariance group H is invariant under the Cartan involution. This
expectation has been borne out in subsequent analyses. However,
the precise coset form of the hidden symmetry group depends upon
performing appropriate dualizations of the fields in the
Lagrangian \cite{lp}. In dimensions nine and above, there is no
enhancement of the $GL(n,{\bf R})$ symmetry. In eight dimensions,
the ${\bf R}_s$ hidden symmetry can be enhanced to an $SL(2,R)$,
and the full global symmetry group takes the form $SL(3,{\bf
R})\times SL(2,{\bf R})$. Likewise, in seven dimensions, ${\bf
G}_s$ takes the form $SL(5,{\bf R})$. Note that, in both cases,
the four-form field strength has been dualized. In six dimensions
and below, there is a potential clash with the target space
duality symmetries of the perturbative string theories \cite{lp},
so let us turn to that subject.

\vskip 0.1in How does the analysis above relate to the appearance
of global symmetry groups in toroidal compactifications of the
supersymmetric string theories? The Ramond-Ramond sector's
antisymmetric $p$-form field strengths must now be distinguished
from the Neveu-Schwarz sector's symmetric and antisymmetric
two-form potentials, $g^{ij}$ and $b^{ij}$, since the latter can
couple directly to the string world-sheet. Thus, if we restrict
ourselves to massless scalars arising as perturbative string
winding or momentum modes, toroidal compactification of either
type II string theory gives the scalar manifold
$O(n,n)/[O(n)\times O(n)]$. Likewise, toroidal compactifications
of the heterotic string give rise to the scalar manifold
$O(n+16,n)/[O(16+n)\times O(n)]$. We must also mod out by the
T-duality group, respectively, $O(n,n;{\rm Z})$, and
$O(n+16,n;{\rm Z})$, which corrects for the over-counting of
equivalent perturbative string compactifications \cite{polbook}:
there is a stringy $R$$\to$$\alpha^{\prime}/R$ symmetry under the
exchange of closed string momentum and winding modes. Finally,
recall that the open and closed type I string theory is not
T-dual, since open strings can't wind. Thus, in this case, the
scalar manifold coincides with what one infers from dimensional
reduction. Notice, as has been emphasized by Lu and Pope
\cite{lp}, that no dualizations of NS-NS sector fields are
necessary in order to make the full T-duality symmetry manifest in
the Lagrangian.\footnote{It should be noted that the coset
structure of the vacuum manifold can equivalently be inferred from
the perspective of the current algebra on the string world-sheet.
Toroidal compactifications are isomorphic to (Lorentzian) even
self-dual lattices \cite{nw,polbook}. The CHL moduli spaces are
supersymmetry-preserving orbifolds of these \cite{cp}.}

\vskip 0.1in The zero slope limit of the massless IIA string is
the same thing as eleven dimensional supergravity compactified on
a circle \cite{cjs,hulltown,witten}--- this is in fact the route
by which the IIA Lagrangian was first constructed \cite{campwest},
and so we expect a correspondence in the global symmetry groups in
dimensions nine and below. Including the $p$ additional axions
obtained by dualizing all Ramond-Ramond sector field strengths,
where $p$$=$$(0,0,0,2,4,8,16,32,64)$ in
$D$$=$$(11,10,9,8,7,6,5,4,3)$ dimensions, the G/H coset takes the
form ${\bf R}^p \times \{ O(n,n;{\rm Z}) \backslash
[O(n,n)/(O(n)\times O(n))]\}$. For $D$$\ge$$7$, the global
symmetry group inferred from perturbative type IIA string theory,
${\bf R}^p \times O(n,n)$, is a subgroup of the Cremmer-Julia
group and there is no clash with T-duality \cite{lp}. But it
should be noted that the dualization of the R-R four-form field
strength was essential in order for this to hold: the T-duality
group is not a proper subgroup of the global symmetry group of the
fully-undualized dimensionally-reduced Lagrangian in $D$$\le$$8$
\cite{lp}. Moreover, in six dimensions and below, it becomes
necessary to also dualize the NS-NS fields in order to enlarge the
manifest global symmetries of the Lagrangian beyond the
perturbative T-duality symmetry group. $GL(11-n,{\bf R})$ and
$O(10-n,10-n)$ are both subgroups of $E_{11-n}$, and their closure
indeed generates $E_{11-n}$. But neither the full ${\bf R}^q$ of
the fully-undualized global symmetry group, nor the full ${\bf
R}^p$ following from full R-R dualization, are contained within
$E_{11-n}$!

\vskip 0.1in To settle this ambiguity, it is incumbent upon us to
understand the significance of alternative dualizations from a
more physical standpoint. Consider the table below summarizing the
results of various possible dualizations which we reproduce from
the paper of Lu and Pope \cite{lp}. We will also note their
observation that the global symmetry group of the
eleven-dimensional supermembrane, $GL(11,{\bf R}) \times R^p$,
contains the perturbative T-duality group of the massless IIA
string, $O(10-n,10-n)$, as a proper subgroup, while $GL(10,{\bf
R})$ does not. Finally, in two dimensions and below, the situation
gets even murkier. It is tempting to continue the Cremmer-Julia
conjecture, arguing for the appearance of the affine extension of
the finite-dimensional Lie algebra $E_8$, namely, $E_9$, in $D=2$,
the first of the hyperbolic Kac-Moody algebras, $E_{10}$, in
$D=1$, and the non-hyperbolic, Lorentzian Kac-Moody algebra,
$E_{11}$, in $D=0$ \cite{west2}. Julia had already conjectured the
appearance of $E_9$ and $E_{10}$ back in 1981 \cite{julia}. But
the entire framework of dimensional reduction and of duality
transformations breaks down in this regime, essentially because a
scalar field can no longer be sensibly distinguished from a gauge
potential in two dimensions.

\bigskip\bigskip
%\vfill\eject

\centerline{
\begin{tabular}{|c|c|c|c|c|}\hline
 &  \multicolumn{3}{c|}{Global Symmetry Groups} & T-duality\\ \hline
$D$& No dualisation & R-R dualisation &Full dualisation & \\
\hline\hline 9 & $GL(2,R)$ & $GL(2,R)$ & $GL(2,R)$ & --\\ \hline 8
& $R*GL(3,R)$ &$SL(3,R)\times SL(2,R)$ & $SL(3,R)\times SL(2,R)$ &
$O(2,2)$\\ \hline 7 & $R^4*GL(4,R)$ & SL(5,R) & $SL(5,R)$ &
$O(3,3)$\\ \hline 6 & $R^{10} * GL(5,R)$ & $R^8 * O(4,4)$ &
$SO(5,5)$ & $O(4,4)$\\ \hline 5 & $R^{20} * GL(6,R)$ & $R^{16} *
O(5,5)$ & $E_{6(6)}$ & $O(5,5)$\\ \hline 4 & $R^{35} * GL(7,R)$ &
$R^{32} * O(6,6)$ & $E_{7(7)}$ & $O(6,6)$\\ \hline 3 & $R^{56} *
GL(8,R)$ & $R^{64} *  O(7,7)$ & $E_{8(8)}$ & $O(7,7)$\\ \hline
\end{tabular}}
\bigskip

\centerline{Table 1: Global Symmetry Groups for Supergravities
with 32 Supercharges in D$\ge$3}
\bigskip\bigskip

\vskip 0.1in It should be noted that there is no difficulty in
correctly identifying the perturbative T-duality group of
compactified string theories in dimensions less than 3.  The
reason is that the world-sheet current algebra and the equivalent
characterization by Lorentzian self-dual lattices, or orbifolds
thereof, continue to be perfectly good tools for identifying the
global symmetry group even when $D$ $\le$ $2$. In fact, both
toroidal, and supersymmetry-preserving orbifold, compactifications
of the heterotic and type II string theories to two dimensions
were widely explored by Chaudhuri and Lowe in \cite{cl}. The basic
message is that it is helpful to shift focus to purely algebraic
techniques in lower dimensions. Indeed, the recent elucidation of
an $E_9$ affine Lie algebra in two dimensions in \cite{nic87,ns}
was based on Nicolai's 1987 reformulation of eleven-dimensional
supergravity, replacing the Lorentz $SO(1,10)$ group with
$SO(1,2)$$\times$$SO(16)$ \cite{nic87}. Note that only an
$SO(1,2)$$\times$$SO(8)$ subgroup of the eleven-dimensional
Lorentz group was preserved in this reformulation: the
Cremmer-Julia symmetry group does {\em not} follow from
straightforward dimensional reduction in dimensions $D$ $\le$ $2$.

\vskip 0.1in Let us now return to the ambiguity in the hidden
symmetry group as a consequence of alternative dualizations in $D$
$\le$ $6$. A hint in the right direction is provided by Roman's
ten-dimensional type IIA cosmological constant \cite{romans},
later identified by Polchinski as the D8brane charge of the type
IIA string theory \cite{dbrane}. We will find that the vexing
problem of accommodating a generator corresponding to a nine-form
gauge potential in the hidden symmetry algebra of M theory leads
to the remarkable conclusion that the Cremmer-Julia $E_{11-n}$
symmetries are not simply one of many options. Rather, they are
{\em required} by the necessity of incorporating both the D8brane
and its magnetic dual. The evidence pointing to this conclusion
comes largely from the interesting recent works of Peter West
\cite{west1,west2,w1,w2,w4}.

\vskip 0.1in We begin with a brief review of work on the M theory
origin of the D8brane, which has been a long-standing puzzle.
Roman's original construction introduced the mass parameter as a
deformation of the field equations of the ten-dimensional IIA
supergravity. Subsequently, Bergshoeff, Green, Hull, Papadopoulos,
and Townsend \cite{berg} showed that, with suitable field
redefinitions, there exists a form of the covariant Lagrangian
with the mass parameter, $M$, appearing explicitly as an auxiliary
field. Taking the limit $M$ $\to$ $0$ smoothly recovers the
massless IIA supergravity Lagrangian. The field equation for the
auxiliary $M$ simply sets the ten-form field strength equal to a
constant $\times$ the epsilon symbol. The explicit appearance of a
nine-form gauge potential in the Lagrangian clarifies how it
couples to an D8brane, but raises the question of its magnetic
dual. Formally, this is a $(-1)$-form gauge potential with scalar
field strength which should couple to a purported D(-2)brane. It
follows that the D8brane is potentially a problem for any
formalism based on the notion of self-duality that makes explicit
use of gauge potentials \cite{witmor,cjlp2,llps}.

\vskip 0.1in For example, a new formalism for supergravity which
extends the coset space description of the scalars to the $p$-form
gauge fields was proposed in \cite{cjlp2} based on doubled fields.
This has the suggestive consequence that the equations of motion
are elegantly formulated as a self-duality condition on the total
field strength, which is a Lie superalgebra-valued object
\cite{julia,cjlp2}. However, the nine-form potential has been left
out of this discussion. When suitably incorporated, as was shown
by \cite{llps}, self-duality forces one to include consideration
of a $(-1)$-form gauge potential. The tension of the associated
D(-2)brane fits neatly into the tower of {\em jade relations}
derived in \cite{llps}: equalities relating the tensions of the
various branes in the duality spectrum. It is natural to ask how
one might sensibly accommodate the notion of a (-1)-form potential
and its associated D(-2)brane.

\vskip 0.1in A partial answer to this question is provided by the
worldsheet formalism of type I$^{\prime}$ string theory. Using the
covariant string theory path integral \cite{flux,cosmo}, we have
shown that the tension of the magnetic dual of the D8brane can be
calculated from first principles. Recall that Polchinski's Dpbrane
tension calculation covered the range $-1$ $\le$ $p$ $\le$ $9$
\cite{dbrane}. The brane-tension was extracted from the
factorization limit of the one-loop open string amplitude with
boundaries on parallel and static Dpbranes: the end points of the
open string lie in a $(p+1)$-dimensional worldvolume. It turns out
that the one-loop amplitude calculated in \cite{dbrane} permits
precisely one possible generalization from the perspective of
two-dimensional Riemannian geometry. This is the one-loop
amplitude with {\em fixed} boundaries, the formalism for which was
developed in the earlier works \cite{cmnp,path}. The factorization
limit of the amplitude with fixed boundaries yields the tension of
an additional Dbrane. Remarkably, the result obtained for the
tension matches perfectly with Polchinski's generic result if we
set $p$ $=$ $-2$. The world-sheet formalism of string theory has
no difficulty accommodating {\em both} a D8brane {\em and} its
magnetic dual, the so-called D(-2)brane, in the duality spectrum.

\vskip 0.1in An elegant and simple explanation of the so-called
(-1)-form potential that lends credence to our result appears in a
work by Schnakenburg and West \cite{w1}. The formalism of doubled
fields \cite{cjlp2} did not clarify how the coset unification of
scalars and gauge fields might be extended to also incorporate the
metric and fermionic fields of supergravity theories. An
alternative approach was subsequently proposed by West
\cite{west1}, in which the entire bosonic sector of both
eleven-dimensional supergravity, and of the massless type IIA and
type IIB supergravities \cite{w0}, were formulated as coset
non-linear realizations of an appropriate Kac-Moody superalgebra.
A key observation made by Cremmer, Julia, Lu, and Pope \cite{cjlp}
highlighted in \cite{west1,west2}, was to note the one-to-one
correspondence between the massless fields of the bosonic sector
of a supergravity theory, and the nodes in the Dynkin diagram of
an $E_{11-n}$ algebra. This leads irrefutably to the conclusion
that the dimensional reduction of eleven-dimensional supergravity
to zero dimensions in the zero volume limit {\em must} result in a
rank 11 Kac-Moody algebra. The question that remains is which
specific algebra.

\vskip 0.1in A hint pointing towards $E_{11}$ is the successful
identification of generators corresponding to the nine-form gauge
potential, {\em and even its magnetic dual}\cite{w1}. The key
observation is that the momentum generator, $P_a$, already plays
the role of the generator corresponding to a putative \lq\lq
$(-1)$-form" gauge potential in the hidden symmetry algebra of the
massive type II supergravity: there is no need to invoke an
additional spacetime field representing the \lq\lq (-1)-form"
generator in the non-linear realization. Thus, the complete
bosonic field content of the massive IIA theory is simply:
\begin{equation}
h_a^b , A, A_c, A_{c_1 c_2} , A_{c_1 c_2 c_3 } , A_{c_1 \cdots
c_5}, A_{c_1 \cdots c_6} , A_{c_1 \cdots c_7} , A_{c_1 \cdots c_8}
, A_{c_1 \cdots c_9} \quad .
\label{eq:IIA}
\end{equation}
$A$ is the dilaton, and $A_{c_1c_2}$ is the Neveu-Schwarz twoform
potential, coupling to perturbative type IIA closed strings. Their
Hodge duals are the eight-form, and six-form, gauge potentials,
respectively. The 1-form and 3-form gauge potentials, and their
Hodge dual 5-form and 7-form potentials, are from the
Ramond-Ramond sector. No dual fields have been introduced for the
nine-form potential, nor for the metric. We now introduce
generators, $R^{c_1 \cdots c_p}$, corresponding to each $p$-form
gauge field listed in Eq.\ (\ref{eq:IIA}). Let us denote the
generators of $GL(10,{\bf R})$ as $K_a^b$, then $GL(10, {\bf R}) $
invariance manifests itself in the commutation relations:
\begin{equation}
[K^a_b , K^c_d ] = \delta_b^c K^a_d - \delta_d^a K^c_b ,
\quad\quad [K^a_b , P_c ] = \delta_c^a P_b , \quad\quad [K^a_b ,
R^{c_1 \cdots c_p } ] = \delta_b^{c_1} R^{ac_2 \cdots c_p} +
\cdots  \quad . \label{eq:gl}
\end{equation}
The difference of the $K^a_b$ are the generators, $J^a_b$, of
$SO(9,1)$ Lorentz transformations. Including the generators of
translations, $P_a$, we have the additional commutation relations:
\begin{equation}
[R, P_a ] =  m b_0 P_a , \quad\quad [P_a , R^{c_1 \cdots c_q } ] =
-m b_q ( \delta_a^{c_1} R^{c_2 \cdots c_q} + \cdots ) \quad .
\label{eq:alge}
\end{equation}
If we now check the commutation relations among the $R^{c_1 \cdots
c_p}$ by themselves, we find that Eq.\ (\ref{eq:alge}) is simply a
special case of this algebra with $P_a$ playing the role of a
putative (-1)-form potential \cite{w1}:
\begin{equation}
[R, R^{c_1 \cdots c_p} ] = c_p R^{c_1 \cdots c_p} , \quad\quad
[R^{c_1 \cdots c_p} , R^{c_1 \cdots c_q} ] = c_{p,q} R^{c_1 \cdots
c_{p+q}}
 \quad . \label{eq:extra}
\end{equation}
We can set $c_{-1} $ $=$ $mb_0$. As pointed out in \cite{w1}, the
limit $m$ $\to $ $0$ smoothly recovers the hidden symmetry algebra
of the massless type IIA supergravity, denoted by ${\cal G}_{\rm
IIA}$ in \cite{w0}. Notice the satisfying agreement with the
physical interpretation of the (-1)-form potential in the
corresponding worldsheet amplitude with fixed boundaries
\cite{flux,path}: the momentum generator acts as a derivative
operator, {\em removing} a worldvolume of codimension one.

\vskip 0.1in The structure constants in Eq.\ (\ref{eq:extra}) can
be determined by verifying consistency with the Jacobi identities
\cite{w1}:
\begin{eqnarray}
c_2 = -c_6 = {{1}\over{2}} . \quad c_3 = c_5 = -{{1}\over{4}} ,
\quad c_1 = - c_{-7} = - {{3}\over{4}}, \quad c_{-1} = -c_9 =
-{{5}\over{4}} \quad \cr c_{1,2} = - c_{2,3} = -c_{3,3} = c_{1,5}
= c_{2,5} = 2 , \quad c_{3,5} = 1, \quad c_{2,6}=2 , \quad c_{1,7}
= 3 \cr c_{2,7} = - 4 , \quad b_2 = - {{1}\over{2}} , \quad b_7 =
- {{1}\over{2}} , \quad b_0 = {{5}\over{8}} \quad .
\label{eq:stra}
\end{eqnarray}
Consistency with the equations of motion of massive type IIA
supergravity, and validity of the proposed non-linear realization
of the hidden symmetry algebra, have been assumed in deriving Eq.\
(\ref{eq:stra}).\footnote{I thank P.\ West for email clarification
of this point.} As was shown by Schnakenburg and West, the
analysis above can be adapted to the type IIB theory \cite{w0},
with algebra denoted as ${\cal G}_{\rm IIB}$. The IIB theory has
doublets of zero-form and two form fields, namely, the dilaton and
axion, and the NS and RR sector two-form potentials with,
respectively, eight-form, and six-form, Hodge dual doublets. No
Hodge dual is introduced for the four-form RR potential, as in
\cite{cjlp2,west1}, since its field strength satisfies a
self-duality condition: the introduction of a \lq\lq double" for a
self-dual field would simply create a mismatch in the number of
physical degrees of freedom. This mismatch is reflected in the
well-known impossibility of writing down a covariant Lagrangian
for the IIB supergravity theory \cite{sw}.\footnote{A clear
explanation of some subtleties in quantizing a self-dual field
strength appears in \cite{witmor}.} The generators of the hidden
symmetry algebra are \cite{w0}:
\begin{equation}
K^a_b , P^a, R_{s}, R_{s}^{c_1 c_2},  R^{c_1 c_2 c_3 c_4},
R_{s}^{c_1 \cdots c_6} , R_{s}^{c_1 \cdots c_8} , \quad \quad s=1
,~ 2 \quad , \label{eq:IIB}
\end{equation}
where $s$$=$$1$, $2$ distinguish potentials in the NS-NS, R-R,
sectors, respectively. The generators satisfy the commutation
relations given in Eq.\ (\ref{eq:gl}), as well as the new
relations:
\begin{equation}
[R_{1}, R_{s}^{c_1 \cdots c_p} ] = d^{s}_p R_{s}^{c_1 \cdots c_p}
, \quad [R_{2}, R_{s_1}^{c_1 \cdots c_p} ] = {\tilde{d}}^{s_2}_p
R_{\epsilon^{s_1 s_2}s_2}^{c_1 \cdots c_p} , \quad [R_{s_1}^{c_1
\cdots c_p} , R_{s_2}^{c_1 \cdots c_q} ] = c^{s_1, s_2}_{p,q}
R_{\epsilon_{s_1 s_2}s_2}^{c_1 \cdots c_{p+q}}
 \quad . \label{eq:extrab}
\end{equation}
Notice that the dilaton and axion differ in their action on the
remaining $p$-forms, respectively, acting so as to preserve, or
switch, the generators within a doublet. Recall that, unlike the
IIA algebra, the spinors of the IIB supergravity have identical
chirality, precluding the possibility of a mass parameter in the
supersymmetry algebra \cite{romans}. Thus, unlike the previous
case, there is no non-trivial extension of the global algebra of
p-form generators by the momentum generator \cite{w0}. The
structure constants of this algebra are determined by requiring
consistency with the Jacobi identities \cite{w0}:
\begin{eqnarray}
d^1_{p+1} = - \quart (p-3) , \quad d^2_{p+1} =  \quart (p-3),
  \quad
{\tilde{d}}^1_2 = - {\tilde{d}}^2_6 = - {\tilde{d}}^2_8 = 1, \quad
{\tilde{d}}^2_2 = {\tilde{d}}^1_6 = {\tilde{d}}^1_8 = 0 \cr
c_{2,2}^{1,2} = - c_{2,2}^{2,1} = -1 , \quad
 c_{2,4}^{2,2} = - c_{2,4}^{2,1} = 4 , \quad c_{2,6}^{1,2} = 1,
 \quad
c_{2,6}^{1,1} = - c_{2,6}^{2,2} = \half  \quad . \label{eq:strb}
\end{eqnarray}

\vskip 0.1in How are the global symmetry algebras of the type II
supergravities related to $E_{11}$ $=$ $E_8^{(3)}$? This question
has been addressed in detail in the very recent paper by West
\cite{w4}, clarifying the precise relationship of ${\cal G}_{\rm
IIA}$, ${\cal G}_{\rm mIIA}$, and ${\cal G}_{\rm IIB}$, to
$E_{8}^{(3)}$. It is remarkable that the global symmetries of each
of the IIA, IIB, and massive IIA supergravities, as well as those
of a broad spectrum of well-known solutions of the classical
supergravities inclusive of the full spectrum of Dbranes and
Mbranes can be elegantly unified within $E_{8}^{(3)}$ \cite{w4}.
In Appendix B, we review West's arguments in more detail, invoking
the framework of nonlinear realizations \cite{cj,cjlp,west2,w4}.

\section{The Theory with Sixteen Supercharges}

\vskip 0.1in In Appendix C, we have reviewed the detailed
prescription for the spacetime reduction of a locally
supersymmetric theory with large $N$ flavor group to a single
point in spacetime given by us in \cite{mtheory}, such that the
resulting zero-dimensional large $N$ matrix model Lagrangian
manifests all of the local symmetries of the original continuum
field theory. Our prescription is a modification of Eguchi and
Kawai's well-known planar reduction procedure, which takes into
account the necessity for an auxilliary local tangent space in a
covariant Lagrangian formulation of a gravitational theory
describing spinors in a generic curved spacetime background. As
our prototype example, we have analyzed the case of the heterotic
10d $N$$=$$1$ supergravity-Yang-Mills Lagrangian with an
anomaly-free Yang-Mills gauge group, $E_8$$\times$$E_8$ or
$SO(32)$, of rank $16$. In part, the reason for this is that a
detailed analysis of the low energy spacetime effective action, up
to quartic order in the inverse string tension, $\alpha^{\prime}$,
and in the inverse Yang-Mills coupling as required by closure
under supersymmetry, exists for the heterotic string supergravity
\cite{br}. This comprehensive analysis is due to Bergshoeff and
Roo \cite{br2,br}, building on the earlier works of
\cite{brwn,ch}. The resulting Lagrangian has been presented in
manifestly supersymmetric form, and in terms of component fields.
The equivalence of the dual two-form and six-form formulations, at
least up to quartic order in $\alpha^{\prime}$, has also been
established by these authors. Partial comparisons have been made,
and are in agreement with, terms in the effective action inferred
from direct string amplitude calculations up to one-loop order
\cite{gsloan}.

\vskip 0.1in Following the c.1995 developments in string duality,
we have an enhanced appreciation of the rich structure of the
vacuum landscape of theories with sixteen supercharges. Toroidal
compactification of the 10d heterotic string preserves all of its
supersymmetries, yielding a rich class of theories with sixteen
supercharges, and anomaly-free rank $16$$+$$n$ Yang-Mills gauge
group, in $10$$-$$n$ spacetime dimensions \cite{nw}. The discovery
of the CHL moduli spaces \cite{chl} clarified that the vacuum
landscape is {\em not} simply-connected: these models are
supersymmetry preserving orbifolds of the standard toroidal
compactifications of the heterotic string \cite{cp}. Thus, for
example, in nine spacetime dimensions the vacuum structure of the
theory with sixteen supercharges is already multiply connected: in
addition to the connected vacuum landscape with 17 abelian
one-forms at generic points in the moduli space, we have an
isolated island universe with 17-8=9 abelian one-forms at generic
points.\footnote{Since the orbifold twist becomes trivial in the
noncompact decompactification limit where the requisite massless
gauge bosons are simultaneously recovered, the CHL orbifold is
not, strictly speaking, a disconnected component of the theory
with 16 supercharges \cite{cp}. But we should emphasize that, at
weak coupling, and in the moduli space approximation, each moduli
space describes low-energy physics in a different island universe.
While nonperturbative dynamics can often be invoked to infer the
possibility of tunneling to a different moduli space with {\em
fewer} supersymmetries \cite{others}, no such examples are known
in theories with 32 or 16 supercharges. Note that the mechanism
proposed for a partial breaking of supersymmetry described in
\cite{frey}, giving a theory with 12 supercharges, requires
assumptions about the nature of the theory in the
decompactification limit.} This theory was first identified as an
asymmetric orbifold of the circle compactification of the
$E_8$$\times$$E_8$ heterotic string theory by Chaudhuri and
Polchinski \cite{cp}: the ${\rm Z}_2$ orbifold action is a
supersymmetry-preserving shift in the one-dimensional momentum
lattice, accompanied by the outer automorphism exchanging the two
$E_8$ lattices. The gauge symmetry at generic points in the moduli
space is rank $9$.

\vskip 0.1in We emphasize that there is no known spacetime
dynamics, field-theoretic or string-theoretic, that can repair the
disconnectedness of the moduli space with sixteen supercharges.
Recall that there is no Higgs mechanism in theories with sixteen
supercharges. Thus, while the precise enhanced gauge group can
vary from point to point, the rank of the abelian subgroup is {\em
fixed} for all points in a connected component of the moduli space
\cite{chl}. More precisely, as is clarified by the orbifold
construction \cite{chl,cp,cl}, each isolated component of the
moduli space is characterized by a distinct target-space duality
group entering into specification of the global symmetry algebra
of that island universe. An alternative viewpoint is to realize
that each island universe is an example of a flux compactification
\cite{ps}: one, or more, of the supergravity pform fluxes is
nontrivial, an invariant on a connected component of the moduli
space. The type IIA string duals of the heterotic CHL models with
nontrivial Ramond-Ramond one-form flux constructed by Chaudhuri
and Lowe \cite{cl} were the earliest known examples of flux
compactifications of the type II string theory. While the notion
of isolated universes can be disconcerting, raising the spectre of
the anthropic principle, and banishing hopes of a unique vacuum
state for String/M theory picked by dynamics alone, we have argued
elsewhere that the problem could be one of misinterpretation
\cite{cosmo}.

\vskip 0.1in Let us move on to a different aspect of the vacuum
landscape of theories with sixteen supercharges, namely, the fact
that the six different string theories: type I, type I$^{\prime}$,
type IIA, type IIB, heterotic $E_8$$\times$$E_8$, and heterotic
$SO(32)$, each describe a different weakly-coupled limit of the
{\em same} moduli space. Consider the circle compactifications of
all six string supergravities and, for convenience, let us
restrict ourselves to discussion of the standard component of the
moduli space characterized by a rank $16$ anomaly-free Yang-Mills
gauge group. As is well-known, the Lagrangian we have described
above can be mapped by a strong-weak coupling duality
transformation, and suitable field identifications, into that of
the type IB string theory \cite{polwit}. Thus, the $SO(32)$ type I
string theory is the strong-coupling dual of the heterotic string
theory with identical gauge group.

\vskip 0.1in In nine dimensions, and below, the $SO(32)$ and
$E_8$$\times$$E_8$ heterotic string theories are related by a
target space duality:
$R_{9}$$\leftrightarrow$$\alpha^{\prime}/R_9$. What is the type I
strong-coupling dual of heterotic vacua with states in the spinor
representations of the orthogonal groups, as required by the
appearance of exceptional Lie algebras? Fortunately, upon
compactification to nine dimensions, the type I theory can acquire
nonabelian gauge symmetries of nonperturbative origin. Under the
T$_9$ duality, the type I string with its 32 D9branes is mapped to
a type I$^{\prime}$ vacuum with 32 D8branes: additional massless
gauge bosons can arise from the zero length limit of D0-D8brane
strings. Such D0-D8brane backgrounds preserve all sixteen
supersymmetries. The incorporation of D0-D8backgrounds, in
addition to those with only 32 D8branes, permits identification of
type I-I$^{\prime}$ strong-weak coupling duals for {\em all} of
the nine-dimensional ground states of the heterotic string
theories. In particular, this includes compactifications on a
circle of both the $Spin(32)/{\rm Z}_2$ and the $E_8$$\times$$E_8$
heterotic string theories \cite{polwit,bachas,flux}. the inclusion
of the D0-D8brane backgrounds also enables the identification of
the type I-I$^{\prime}$ strong coupling duals for all of the
heterotic CHL moduli spaces with sixteen supersymmetries
\cite{chl,cp,cl}. Furthermore, since type I$^{\prime}$ theory
compactified on $S^1$ is the same thing as M theory compactified
on $S^1$$\times$$S^1/{\rm Z}_2$, these observations are consistent
with the identification of M theory on $S^1/{\rm Z}_2$ as the
strong coupling limit of the $E_8$$\times$$E_8$ heterotic string
theory \cite{hulltown,hw}.

\vskip 0.1in Most importantly, the successes in unifying the
circle-compactified heterotic-typeI-I$^{\prime}$ theories with
sixteen supercharges can be extended to incorporate the type IIA
and type IIB string theories. There exists a nine-dimensional
Lagrangian formulation of the massive type IIA-IIB string theories
due to Bergshoeff, de Roo, Green, Papadopoulos, and Townsend
\cite{berg} which incorporates the full spectrum of Dbrane
$p$-form potentials \cite{dbrane}, including Roman's IIA
cosmological constant \cite{romans}. By combining
field-dualizations, as well as $S$ and $T$-duality transformations
on the couplings, this Lagrangian can be mapped to {\em any} of
six supergravity theories: the circle-compactified type I, type
IIA, or heterotic string supergravities, the Scherck-Schwarz
reduction of the type IIB string supergravity, or the
$S^1$$\times$$S^1/{\rm Z}_2$ compactification of
eleven-dimensional supergravity. This covers all six vertices of a
modified star diagram linking theories with sixteen supercharges
\cite{hulltown,polbook,mat1}.

\vskip 0.1in With our new prescription for spacetime reduction, we
have shown that the local symmetries of the supergravity
Lagrangian can be preserved in the reduced matrix model. Thus,
there is a precise analog for each field redefinition, or
dualization, of the continuum Lagrangian in the matrix model: the
matrix Lagrangian is only unique upto appropriate dualizations
defined on the matrix variables \cite{mat1}. On the one hand, this
is a beautiful illustration of the unity of the different string
theories with eleven-dimensional supergravity. But it points to
the importance of understanding the global symmetry algebra: the
identification of a specific, hidden symmetry algebra is what
gives precise meaning to one, or other, class of supergravity/M
theory toroidal compactifications. This observation has been
reiterated recently by West \cite{w2,w1,w4,w5}. But it is not new
to string theory, nor to supergravity: target-space duality
groups, and their conjectured extension to U-dualities
\cite{hulltown}, have been the bulwark of our understanding of
string and M theory compactifications. In section 5 of this paper,
we will argue that the notion of the global symmetry algebra also
provides a precise generalization incorporating {\em all} of the
ground states of M theory, beyond toroidal compactifications.
Remarkably, this will include the isolated island universes
discovered in \cite{chl}.

\vskip 0.1in We began this section by pointing out that, at the
current time, we do not have a comprehensive analysis of the full
covariant Lagrangian--- including all of the fermionic
contributions necessitated by supersymmetry, for any of the low
energy effective Lagrangians other than that of the heterotic
string theory \cite{br2,br}. We save this for future work. The
important puzzle of how one might incorporate theories with 32
supercharges, and no Yang-Mills fields, as a special limit of the
vacuum landscape of theories with sixteen supercharges will be
addressed in the next section.

\section{M Theory and its Hidden Symmetry Algebra}

\vskip 0.1in We have alluded earlier to the existence of a hidden
symmetry algebra in the matrix Lagrangian that is larger than the
obvious $U(N)$$\times$$G$. In part, there is an $SL(10,{\bf R})$
symmetry, which is the manifest remnant under spacetime reduction
to a single point in spacetime of the Lorentz symmetry group of
the 10d continuum field theory Lagrangian. In the Introduction, we
have already explained the simple rationale for expecting the
symmetry algebra of eleven-dimensional supergravity, and
consequently, of M theory, to be $E_{11}$ $=$ $E_8^{(3)}$, the
rank eleven algebra known as the very-extension of the finite
dimensional Lie algebra $E_8$ \cite{w4}. How does the symmetry
algebra of the theory of sixteen supercharges relate to that of M
theory? We address that question in this section, following the
discussion given in \cite{mtheory}.

\vskip 0.1in The nonmaximal 10d supergravity is a theory with
sixteen supercharges. In the notation of \cite{w4}, we have the
following symmetry generators:
\begin{equation}
K^a_b , R , R^{c_1 c_2} , R^{c_1 \cdots c_6} , R^{c_1 \cdots c_8}
 \quad .
\label{eq:typei}
\end{equation}
The heterotic supergravity theory has zero-form dilaton and NS
two-form potentials, plus their ten-dimensional Hodge duals,
respectively, six-form and eight-form supergravity potentials. The
$K^a_b$ are the generators of $GL(10,{\bf R})$ in the notation of
\cite{w4}. The commutator algebra of these generators was given in
\cite{w5}. Not surprisingly, we will find that it agrees precisely
with the algebra that can be inferred from the appropriate
projection on the global symmetry algebra of the type IIB
supergravity. This reflects the well-known connection between
these two string theories under the orientation projection
identifying left-moving and right-moving modes on the worldsheet
\cite{polbook}.

\vskip 0.1in The global symmetry algebra of the type I
supergravity can be identified by taking an appropriate projection
of the global symmetry algebra, ${\cal G}_{IIB}$, of the 10d type
IIB supergravity, which has been obtained in a recent work of
Schnakenburg and West \cite{w2}. By setting the extra forms to
zero in Eqs.\ (1.1-.3) of \cite{w2}, we find the usual $GL(10,{\bf
R})$ algebra, extended by translations:
\begin{equation}
[K^a_b , K^c_d ] = \delta_b^c K^a_d - \delta_d^a K^c_b ,
\quad\quad [K^a_b , P_c ] = \delta_c^a P_b , \quad\quad [K^a_b ,
R^{c_1 \cdots c_p } ] = \delta_b^{c_1} R^{ac_2 \cdots c_p} +
\cdots  \quad , \label{eq:gl10}
\end{equation}
plus the simplified algebra of 0, 2, 6, and 8-form generators:
\begin{equation}
[R , R^{c_1 \cdots c_p} ] = d_p R^{c_1 \cdots c_p}  , \quad
[R^{c_1 \cdots c_p} , R^{c_1 \cdots c_q} ] = c_{p,q} R^{c_1 \cdots
c_{p+q}}
 \quad . \label{eq:extrabI}
\end{equation}
Comparing with the IIB result given in Eq.\ (1.3) of \cite{w2},
the remnant non-vanishing structure constants take the simple
form:
\begin{equation}
d_{q+1} = - \quart (q-3) ,  ~ q=1,5, \quad\quad \quad c_{2,6} =
\half \quad . \label{eq:strbI}
\end{equation}
The algebra we obtain is in precise agreement with that of the 10d
$N$$=$$1$ heterotic supergravity given in Eq.\ (1.4) of \cite{w5}.
Let us denote this algebra as ${\cal G}_{IB}$. We emphasize that,
thus far, we have not included the Yang-Mills gauge sector of the
nonmaximal 10d supergravity-Yang-Mills theory.

\vskip 0.1in Let us now address the question of how the global
symmetry algebra of the heterotic-type I nonmaximal supergravity
relates to the symmetry algebra of M theory. West has provided
mounting evidence in favor of the conjecture that the symmetry
algebra of M theory is the rank eleven very-extended algebra
$E_{11}$ \cite{w4}, a summary appears in Appendix B. If we compare
the generators and commutation rules given above with the
Chevalley basis for the algebra $E_8^{(3)}$, written in either its
IIA or IIB guise as shown in \cite{w4}, we find that we are
missing some of the positive root generators in either
formulation. We have all of the generators, $E_a$$=$$K^a_{a+1}$,
$a$$=$$1$, $\cdots$, $9$, of $SL(10,{\bf R})$. In the IIA
formulation, given in Eq.\ (4.4) of \cite{w4}, we are missing the
roots corresponding to the R-R one-form, and NS-NS twoform,
namely, $E_{10}$$=$$R^{10}$, and $E_{11}$$=$$R^{910}$. In the IIB
formulation, we are missing the roots labelled
$E_{9}$$=$$R_1^{910}$, and $E_{10}$$=$$R_2$, arising,
respectively, from the NS-NS two-form potential, and R-R scalar.
It is clear we cannot build a full $E_{8}^{(3)}$ algebra from the
restricted set of generators in ${\cal G}_{IB}$.

\vskip 0.1in In \cite{w5}, it was pointed out that a different
rank eleven very-extended algebra, namely, the very-extension of
the $D_8$ subalgebra of $E_8$, can be spanned by the generators of
${\cal G}_{IB}$. But we should note that such a construction is
somewhat unmotivated from the viewpoint of any relationship to the
type II theories, to eleven-dimensional supergravity, or to M
theory: the authors of \cite{w5} make the choice
$E_a$$=$$K^a_{a+1}$, $a$$=$$1$, $\cdots$, $9$, $E_{10}
$$=$$R^{910}$, and $E_{11}$$=$$R^{5678910}$. This choice of simple roots
is shown to generate the very-extended algebra $D_8^{(3)}$.
Appending a one-form generator to this set converts the
$D_8^{(3)}$ algebra to $B_8^{(3)}$ \cite{w5}. However, it should
be noted that, since the two-form and six-form potentials are
Hodge dual to each other in ten dimensions, a construction which
includes both in the simple root basis is quite different in
spirit from that of the $E_{8}^{(3)}$ algebras underlying the IIA
and IIB theories \cite{w4}.

\vskip 0.1in We will suggest a different direction towards
uncovering an $E_8^{(3)}$ algebraic structure in supergravity
theories with sixteen supercharges. Since we already have the
requisite two-form potential, respectively, the R-R, or NS-NS,
two-form of the type IB, or heterotic, supergravities, our goal
will be to identify a one-form potential that can play the role of
the positive root generator labelled $E_{10}$ in the IIA
formulation of the $E_8^{(3)}$ algebra. A hint is provided by our
understanding of the duality web linking the zero slope limits of
the circle compactifications of the type I, type IIA, IIB, and
heterotic string theories, along with M theory compactified on
$S^1$$\times$$S^1/{\rm Z}_2$. Upon compactification on a circle,
the heterotic string theories acquire an abelian one-form
potential, namely, a Kaluza-Klein gauge boson. This perturbative
gauge field is known to play a crucial role in six-dimensional
weak-strong IIA-heterotic string-string duality: it maps under a
weak-strong coupling duality to the R-R one-form potential of the
IIA string theory compactified on K3.\footnote{To be more precise,
compactifications of the IIA theory on K3 are described by a
$(19,3)$ cohomology lattice characterizing classical K3 surfaces.
It is the quantum extension to a $(20,4)$ quantum cohomology
lattice, as a consequence of introducing a flux for the R-R
one-form potential, that completes the isomorphism of the IIA
theory compactified on K3 to the heterotic string compactified on
$T^4$. The quantum cohomology lattice is identified with the
$(20,4)$ Lorentzian momentum lattice of the heterotic string
\cite{sen1,harvs,cl}. The heterotic theory has 20 abelian
one-forms. One of these is distinguished as the partner of the R-R
one-form of the IIA theory, and this is the Kaluza-Klein gauge
field we have in mind.} As the simple root generator labelled
$E_{11}$ in the IIA formulation of $E_8^{(3)}$, we choose,
respectively, the NS-NS two-form potential of the IIA string or
the two-form potential of the heterotic string. Note that these
are mapped to each other under string-string duality. As an aside,
the reader may wonder why we had to compactify all the way to six
dimensions to see these equivalences, but this methodology is in
keeping with how the Cremmer-Julia hidden symmetries of the type
II theories were discovered. The full global symmetries only
become manifest in the dimensionally-reduced supergravity
theories, but this can be taken as a hint towards discovering a
higher-dimensional correspondence. In summary, identifying the
Kaluza-Klein one-form, and the two-form potential, as the two
missing positive root generators in the Chevalley basis should
plausibly allow one to demonstrate an $E_8^{(3)}$ global symmetry
algebra in the circle compactifications of the heterotic
supergravities.

\vskip 0.1in The evidence for an $E_8^{(3)}$ global symmetry
algebra is even clearer in the case of the circle-compactified
type IB supergravity. Upon compactification to nine dimensions,
the type I theory can acquire nonabelian gauge symmetries of
nonperturbative origin. The type I$^{\prime}$ theory has the same
$p$-form gauge potentials as the massive IIA theory, and so the
correspondence to the $E_8^{(3)}$ hidden symmetry algebra is
particularly transparent. It is important to notice that the
supergravity structure of the CHL orbifold is identical with that
of the circle compactification. Indeed, if we are correct in our
expectation that the circle-compactified theory has an underlying
$E_8^{(3)}$ global symmetry algebra, then this will also be true
of the CHL orbifold. The distinction between the two theories lies
in the Yang-Mills sector: they differ in the rank of the gauge
group at generic points in the moduli space, respectively, 17, and
9. Are the precise nonabelian enhanced gauge symmetry groups
relevant to this discussion? It has become customary to think not,
since it is well-known that the precise nonabelian enhancement
varies from point to point in the moduli space. Conventionally,
this multiplicity of enhanced symmetry points is expressed in
terms of the perturbative T-duality groups of the moduli spaces.

\vskip 0.1in However, if the Kaluza-Klein one-form is absorbed in
the nonlinear realization of $E_8^{(3)}$, an especially simple
result follows: we find three distinct theories in nine spacetime
dimensions with sixteen supercharges. They are characterized by
three distinct global symmetry groups: $E_8^{(3)}$$\times$$({\rm
Spin}(32)/{\rm Z}_2)$, $E_8^{(3)}$$\times$$(E_8$$\times$$E_8)$,
and $E_8^{(3)}$$\times$$(E_8)$.\footnote{We are using the same
notation for the Yang-Mills gauge group and its global remnant.
The $E_8$ in the case of the CHL orbifold is realized at level two
of the worldsheet affine Lie algebra. From the perspective of
spacetime, the ${\rm Z}_2$ orbifold structure indicates the
necessity for care in determining the precise global remnant of
the $E_8$ gauge symmetry.} The factor in brackets arises from the
Yang-Mills sector, the former from the supergravity sector. The
strong coupling duals of these heterotic theories are,
respectively, the circle compactified type IB, type I$^{\prime}$,
and type I$^{\prime}$, theories in nine spacetime dimensions with
corresponding global symmetry groups.

\vskip 0.1in From the perspective of the hidden symmetry algebra,
on the other hand, there is hope of finding a tantalizing
relationship between supergravity theories having 16 or 32
supercharges: M theory would have a superalgebra given by the
fermionic extension of $E_8^{(3)}$ by 32 supersymmetry generators.
An alternative extension of the bosonic algebra $E_8^{(3)}$ to a
superalgebra with only sixteen supersymmetry generators would
simultaneously permit the incorporation of up to 32 Yang-Mills
gauge fields. More precisely, we would find that the possible
extensions of the global symmetry group are isomorphic to the
isolated components of the moduli spaces with sixteen
supercharges. In nine dimensions, the only solutions are the rank
16 and rank 8 groups described above. But for moduli spaces in
lower dimensions, there is a huge proliferation of isolated
components to the moduli space, with the enhanced gauge symmetry
varying from point to point, but with identical abelian subgroup
at all points in the moduli space. Thus, a precise specification
of the global remnant of this abelian subgroup is an accurate
characterization of the CHL orbifold. Notice that the nontrivial
global remnant of the abelian gauge symmetry simply reflects the
action of the orbifold group on the toroidal spacetime. This is a
slightly more physical explanation of how physics differs in the
CHL moduli spaces.

\vskip 0.1in In summary, let us reformulate our conjecture for M
Theory succinctly \cite{mat1,mtheory} in light of what we have
learned from this brief survey of the hidden symmetry algebras of
nine, ten, and eleven, dimensional supergravity theories. Given
the striking evidence that the rank eleven very-extension of the
Lie algebra $E_8$ incorporates the full spectrum of NS-NS and R-R
pbranes, $-2$$\le$$p$$\le$$9$, we will conjecture that the hidden
symmetry algebra of supergravity theories with sixteen
supercharges takes the form $E_8^{(3)}$$\times$$ G$, where $G$ is
the global remnant of an appropriate finite-dimensional Yang-Mills
gauge group. We conjecture further that nonperturbative String/M
theory is the locally supersymmetric extension of the algebra
$E_8^{(3)}$$\times$$G$ with sixteen supercharges realized on the
field of unitary $N$$\times$$N$ matrices. A particular limit of
this superalgebra will recover a full 32 supercharges at the cost
of making $G$ trivial.

\section{Open Questions}

\vskip 0.1in A proposal as radical as that described in
\cite{mat1,mtheory} has few concrete conclusions in comparison
with the Pandora's box of fascinating questions it opens up for
future investigation. We include the list of the most significant,
and accessible, of the questions given in \cite{mtheory} below,
along with some issues specific to the framework of nonlinear
realizations and Lorentzian Kac-Moody algebras
\cite{w4,dhn,eh,ganor,nk}:

\begin{itemize}
\item{What is the physical significance of the higher level roots
of the Lorentzian very-extended Lie algebra? This is the most
important issue in extending the arguments {\em in favor} of
$E_{11}$ as the symmetry algebra of M theory into a fully
convincing {\em proof}. All Lorentzian Kac-Moody algebras exhibit
the level structure, described briefly in Appendix A, and it is
here that $E_{11}$ has consequences for M theory beyond those
evident in its low energy supergravity limit. Fortunately, there
is considerable ongoing investigation of the level structure, both
in the context of the over-extended algebra $E_{10}$$=$$E_8^{(2)}$
\cite{ganor,dhn,nk}, and in the case of $E_{11}$$=$$E_8^{(3)}$
\cite{ksw,kw,eh,w4}. The general feature common to all of these
analyses is to attempt an isomorphism between all of the known
classical supergravity solutions, including composite branes and
bound states, to the root system of the Lie algebra. There will
also be new solutions which do not have a supergravity
correspondence; examples in the case of eleven-dimensional
supergravity have been given by West in \cite{w4}. An interesting
point raised by West \cite{w4} that deserves further investigation
is the possibility of an isomorphism between the process of group
multiplication between group elements corresponding to two
elementary branes, and the formation of a composite brane or bound
state. How should one interpret such states, and what is their
correspondence in the matrix theory framework?}

\item{The absence of a clear picture for the origin of spacetime
in proposals for nonlinear realizations of the hidden symmetry
algebra \cite{ganor,dhn,eh,west2,w4} has made it difficult thus
far to explore their relationship to more traditional organizing
principles for the pbrane spectrum of type II string/M theory,
such as K-theory \cite{witk,dmw}.\footnote{I thank Clifford
Johnson for raising this question.} We should note that the
precise role of self-duality in the context of K theory is also
unclear. Preliminary steps could be to understand the relationship
of K theory to the doubled field formalism of
dimensionally-reduced supergravities \cite{cjlp2}, a precursor to
West's nonlinear realization \cite{west2,w4}, also based on the
notion of self-duality \cite{witmor}. Given the recent detailed
understanding provided by West in \cite{w4} of the full pbrane
spectrum of the type II string theory, M theory, and of
eleven-dimensional supergravity, it must surely be possible to
make contact with at least some of the results in \cite{dmw}.
Recent work in this direction has exploited the worldsheet
correspondence, invoking the framework of RG flows in the larger
space of generic two-dimensional field theories, thus
incorporating unstable branes. Do unstable branes find a natural
setting within the framework of nonlinear realizations?}

\item{The elucidation of an $E_8^{(3)}$ algebra with Chevalley
basis chosen from among the generators of the global symmetry
algebra of the circle-compactified type I-I$^{\prime}$-heterotic
string theories proposed by us in \cite{mtheory}, and reviewed in
section 4, needs to be completed. In particular, West has made the
interesting observation that the IIA-IIB T-duality transformation
simply reflects the bifurcation symmetry of the $E_8^{(3)}$ Dynkin
diagram at its central node, interchanging the Dynkin diagrams of
its two inequivalent $A_{9}$ subalgebras \cite{w4}. This argument
can clearly be adapted to the T-duality symmetry relating the type
I and type I' string theories. Or, to that relating the two
circle-compactified heterotic string theories. The details need to
be verified.}

\item{As a further check, the conjectured hidden symmetry algebra,
$E_8^{(3)}$$\times$$E_8$, for the first of the CHL models
\cite{cp} needs to be verified. In fact, we have conjectured the
appearance of an $E_8^{(3)}$ algebra in the supergravity sector of
{\em any} of the vast proliferation of isolated theories with
sixteen supercharges \cite{others}. It should be noted that our
new perspective on the hidden symmetry algebra takes seriously the
strong-weak dualities linking the heterotic, type IB, and type
IIA, string theories: a theory of sixteen supercharges is
self-dual, and it would be meaningless to have different hidden
symmetry algebras pertaining to the different string theories.
Thus, while the perturbative T-duality group of the type II theory
is, in fact, incorporated in $E_{11}$, the Lorentzian $O(16+d,d)$,
$d <10$, extension familiar from the toroidally-compactified
heterotic string is {\em not} contained within $E_{11}$. Our
conjecture is that $E_{11}$ is the hidden symmetry algebra of the
type I-heterotic supergravity with sixteen supercharges: it is
unchanged for {\em all} of the CHL theories. Of course, the
abelian subgroup of the nonabelian gauge symmetry characterizing
generic points in a given moduli space, and hence $G$, will be
different for each of the latter CHL theories. It just so happens
that in nine dimensions there are no Wilson lines that permit
either a breaking, or enhancement, of the $E_8$ Yang-Mills gauge
symmetry.\footnote{I thank Arjan Keurentjes for requesting this
clarification.}}

\item{Next, we must ask about theories with fewer supercharges:
note that the field theory with eight supercharges includes in its
moduli space both subspaces, or isolated points, with 4, or 0,
supersymmetries, at finite distances in the moduli space. The
appearance of matter fields is an added wrinkle. Can the framework
of nonlinear realizations be adapted to supergravity theories
coupled to chiral matter? This is a beautiful open question that
first needs to be addressed purely in a field theoretic setting.
Our conjecture is that the theory with eight, or fewer,
supercharges has a hidden symmetry algebra that is {\em smaller}
than $E_8^{(3)}$, possibly a subalgebra. The N=3 theories with 12
supercharges, recently revived by Frey and Polchinski \cite{frey},
offer an interesting half-way point between the 16, 32 supercharge
theories and theories with 8 or fewer supercharges.\footnote{I
thank Andrew Frey for stimulating my interest in this subject.}
They have well-defined moduli spaces with flat potential, while
including examples of three-form flux compactifications that share
many features of semi-realistic, N=1 flux compactifications with
avenues for moduli stabilization. If a suitable generalization of
the framework of nonlinear realizations and hidden symmetries can
indeed be found for such semi-realistic supergravity theories, it
should be straightforward to implement within the context of
matrix theory.}

\item{Once the precise nature of the hidden symmetry algebra of
the type I-I$^{\prime}$-heterotic theories with sixteen
supercharges has been pinned down, and which we have conjectured
will take the form $E_8^{(3)}$$\times$$G$, what new physics
becomes accessible? It is a remarkable fact that there exists a
{\em unique} assignment of phases in the bosonic $E_8^{(3)}$
algebra corresponding to an eleven-dimensional theory with,
respectively, Minkowskian (1,10), or Euclidean (0,11) signature
\cite{keurentjes}. As has been shown by Keurentjes
\cite{keurentjes}, every other self-consistent choice of phases
for $E_{11}$ results in a spacetime with two, or more, timelike
directions. Furthermore, the Euclidean case also corresponds to a
bosonic $E_{11}$ algebra without supersymmetric extension
\cite{keurentjes}, precisely as one would expect given its natural
physical interpretation as the symmetry algebra of M theory at
finite temperature. This Euclidean symmetry algebra should be of
great interest in the context of M theory cosmology, as we now
explain.

\vskip 0.1in We will offer a suggestive interpretation for the
principle subalgebra of the Euclidean signature bosonic $E_{11}$
algebra. As expected from generic considerations \cite{nico},
every Lorentzian Kac-Moody algebra has a principal $SO(1,2)$
subalgebra, and it is natural to seek its physical interpretation.
Based on our understanding of the String/M duality web in 11, 10,
and 9, spacetime dimensions, and given the pivotal role played by
the nine-form potential and its (-1)-form dual, it is natural to
identify the parameters of the $SO(1,2)$ subalgebra, roughly, as
follows.\footnote{We should emphasize that this is very rough
intuition. Thus, we are not making any clear statement on the
precise topology of the group, discrete identifications might be
necessitated.} Labelling them as $R_0$, $R_{10}$, and $R_9$,
respectively, suggests a natural identification with inverse
temperature, $\beta$, string coupling, $g$, and cosmological
constant $M$. The latter is Roman's mass parameter, later
interpreted by Polchinski as D8brane charge \cite{romans,dbrane}.
Notice that the two-parameter subspace $(\beta, g)$, whose rough
correspondence with the radius of coordinates $(X^0,X^{10})$ is
well-known, is also the classic phase space parameterization
relevant for the study of the dynamics of a finite temperature
gauge theory \cite{decon}. Supplementing this with the
cosmological constant gives a natural three-parameter phase space
relevant for discussions of cosmology: the thermal dynamics of the
Universe, inclusive of gravity \cite{decon,cosmo}. It should be
emphasized that the principal $SO(1,2)$ algebra should {\em not}
be confused with the corresponding subalgebra of the spacetime
Lorentz algebra.

\vskip 0.1in In recent work on string thermodynamics \cite{decon},
we have pointed out that there is a fundamental conceptual barrier
to proposals for a {\em microcanonical} description of the string
ensemble: perturbative string theory is inherently a
background-dependent theory. Thus, we cannot escape the \lq\lq
heat-bath" represented by the spacetime geometry, and additional
background fields: {\em any} self-consistent discussion of string
thermodynamics must therefore be relegated to the canonical
ensemble. Fortunately, there is a first-principles framework for
the canonical ensemble provided by the world-sheet path integral
formalism, originally pointed out in \cite{poltorus}. On the other
hand, from the perspective of quantum cosmology, the ensemble of
interest {\em is} the microcanonical ensemble of the fundamental
degrees of freedom: the Universe is a closed system, and there is
no \lq\lq heat bath" one can point to. The reduced matrix models
we have described in this paper offer a self-consistent starting
point in which to formulate the microcanonical ensemble of the
fundamental degrees of freedom. This opens up the exciting
possibility of a genuinely nonperturbative formulation for black
hole thermodynamics and quantum cosmology.}

\item{Next, the detailed understanding of the supersymmetric
extension of the $E_8^{(3)}$$\times$$G$ hidden symmetry algebra is
of profound importance, especially in light of our conjecture for
the appearance of the theories with 32 supercharges, and no
Yang-Mills sector, as special limits in the superalgebra with
sixteen supercharges. Fortunately, in recent work \cite{nk}, some
of the tools necessary for such an analysis have been developed.
We should note, however, that the focus of \cite{nk} is the rank
10 hyperbolic Kac-Moody algebra $E_{10}$. Recall that the $A_{10}$
subalgebra of $E_{11}$ has the natural interpretation as
originating in the gravity sector of a {\em noncompact} 11d
theory. $E_{10}$ has an $A_9$ subalgebra, but not $A_{10}$. Not
surprisingly, the framework of \cite{nk} also cannot incorporate a
space-filling nine-brane, which would have to couple to a rank
eleven field strength.}

\item{Given the precise details of the hidden {\em super}algebra,
an analysis along the lines of \cite{brwn,br2,br} is necessary to
verify whether there exists a manifestly supersymmetric and
covariant supergravity and Yang-Mills Lagrangian where this
symmetry is manifest? If so, by the arguments in this paper, it is
clear that there exists a corresponding zero-dimensional
supermatrix Lagrangian where this symmetry is also manifest. We
should emphasize that whether one implements our nnperturbative
proposal for M theory in a Lagrangian or Hamiltonian framework is,
in part, a matter of taste. Each perspective offers distinct
advantages. Nevertheless, for the reasons mentioned above, the
matrix Lagrangian formulation should remain a priority for future
investigations.}

\item{As an aside, we should note that an important issue raised
both by our focus on the principal three-parameter subalgebra of
$E_{8}^{(3)}$, and by its realization in a locally supersymmetric
unitary matrix model, is the possibility of an undiscovered
relation to the famous {\em supermembrane theory}, a conjectured
theory of fundamental supermembranes \cite{sm,dhopn}. Does the
locally supersymmetric matrix model represent a regularization of
the three parameter manifold of the principal subgroup, analogous
to the regularization of the worldvolume of the supermembrane
provided by the rigid unitary matrix model \cite{dhopn}? These are
open questions that should shed light on the large N continuum
limit of the reduced matrix model.}

\item{Finally, coming to the crucial open questions in the matrix
model framework, there is the issue of what comes beyond leading
order in large $N$ in the matrix model: what is the significance
of the off-diagonal elements of the variables in the reduced
matrix model? Notice that there is an obvious extension to the
notion of the double-scaling limit familiar from the $c=1$ matrix
model: namely, $\lim_{N \to \infty , g \to 0}$, with $g^{\alpha}
N^{\beta}$ held fixed. The parameters $(\alpha, \beta)$ take an
appropriate range of values for members in the discrete series of
the gravitationally-dressed unitary conformal field theories with
central charge $c$$\le$$1$ \cite{mm}. The generalization to large
$N$ limits with multiple-scaling was pointed out in our earlier
works \cite{mat1}. Since we have a full range of background
fields, $(g=e^{\bar{\phi}}, {\bar{A}}_{c_1}, {\bar{C}}_{c_1},
{\bar{C}}_{c_1c_2}, \cdots , {\bar{C}}_{c_1 \cdots c_9})$, where $
g$$=$$(M_{11}R_{10})^{3/2}$, and the single mass scale, $M_{11}$,
there are many possible inequivalent, multiple-scaling limits: a
suitable combination of powers of $N$, $M_{11}$, and the
background fields, can be held fixed, in the limit that we take
$N$ $\to$ $\infty$. Here, $M_{11}$ has been taken to be the
eleven-dimensional Planck mass. The precise powers of $M_{11}$
that enter into taking the large $N$ limit can vary, depending on
whether we wish to match to an eleven, or ten-dimensional,
continuum field theory. For example, the ten-dimensional string
mass scale is related as follows: $m_s$ $=$ $\alpha^{\prime -1/2}$
$=$ $M_{11}^{3/2} R_{10}^{1/2}$. We should emphasize the fact that
it was essential that the matrix Lagrangian framework allow for a
wide range of inequivalent large $N$ limits, since it would not
otherwise be possible to explain the multitude of known effective
dualities relating M theory ground states.\footnote{As an
illustration, consider the unusual suggestion for incorporating
nonsupersymmetric, metastable ground states such as the heterotic
orbifold with {\em timelike} linear dilaton potential and matter
fields with $D>1$, recently explored in \cite{simeon}. The
relevant scaling limit would requires matching to a matrix model
with extended symmetry algebra embedding $SL(10+n,{\bf
R})$$\times$$SO(32+n)$$\times$$SO(n)$, where $n$ is the number of
extra matter bosons, and with nontrivial background dilaton and
three form flux. Note that the generators of $SL(10+n,{\bf R})$
can nevertheless be rescrambled into generators of $E_8^{(3)}$,
except that some of them will now appear at higher level in the
Lorentzian algebra, as discussed from a rather general viewpoint
in \cite{w4}. I thank Simeon Hellerman for discussion of string
theory backgrounds with timelike dilaton potential.}}

\item{Perhaps the most important open question is the comparison
of corrections to the large $N$ limit of the matrix model,
calculated with the specific choice of scaling appropriate for
matching to a particular string supergravity, with the higher
order in $\alpha^{\prime}$ corrections to the string spacetime
effective Lagrangian.\footnote{I would like to thank Sanefumi
Moriyama for asking this question.} We should remind the reader
that the precise form of the low energy spacetime effective
Lagrangian for string theory has not been systematically
calculated beyond quartic order in the inverse string tension
\cite{br2,br}, and that too only in the case of the heterotic
string. This is unfortunate, given that the techniques for the
systematic derivation of these terms from string amplitude
calculations, or based on duality symmetries of the effective
action, have been known for many years \cite{br2,gsloan,polbook}.
In the past, this was explained by the absence of any direct
physical motivation for a {\em comprehensive} analysis. For
example, it was common to focus on the particular subset of terms
that had the potential to mediate some new physics beyond the
standard model. But given our current understanding of M theory,
it would now seem that there is {\em strong motivation} for a
renewed effort at obtaining a comprehensive analysis of the
spacetime effective Lagrangian. We emphasize that it is only at
higher orders in the $\alpha^{\prime}$ expansion that we can
successfully test any conjectured nonperturbative proposal for M
theory beyond agreement with the supergravity prediction.}

\end{itemize}

\noindent In summary, we believe this could be the beginning of an
exciting period in the search for a more fundamental description
of String/M theory that transcends its weakly-coupled perturbative
limits.

\vskip 0.2in \noindent{\bf ACKNOWLEDGMENTS} \vskip 0.1in I would
like to thank Bernard de Wit, Thibault Damour, Andrew Frey, Ori
Ganor, Jaume Gomis, Clifford Johnson, Simeon Hellerman, Arjan
Keurentjes, Marco Matone, Andrei Mikhailov, Sanefumi Moriyama,
Hermann Nicolai, Igor Schnakenburg, and Peter West for helpful
comments and stimulus. I thank Andrew Frey for the opportunity to
present this material to a Caltech seminar audience. The paper was
written in part at the Kavli Institute for Theoretical Physics,
University of California at Santa Barbara. I am grateful to the
staff for providing a supportive and stimulating environment for
research on an informal visit.

\vskip 0.2in \appendix \noindent{\large{\bf Appendix A: The
Very-Extension of a Lie Algebra}}

\vskip 0.1in  The introductory material in this appendix can be
found in the classic texts \cite{kac,waki}, as well as the
well-known review article \cite{go}. The notion of a very-extended
algebra first appears in Gaberdiel, Olive, and West \cite{w2}. Our
discussion of the very-extension of $E_8$ is based on section 5
and the appendices of this reference. We urge the reader to
consult the original \cite{w2} for a far more complete treatment.
A readable introduction to some novelties in the representation
theory of very-extended algebras, the subject of current research,
can be found in section 2 of \cite{kw}.

 \vskip 0.1in As is well-known, Cartan's classification of the
classical Lie algebras, the $A_n$, $B_n$, $C_n$, and $D_n$ series,
extends to three exceptional cases, namely, $E_n$, with $n$$=$$6$,
$7$, and $8$ \cite{gilmore}. Every such finite-dimensional
semi-simple Lie algebra has an infinite-dimensional affine
extension, better known in the physics literature as a current
algebra \cite{ccwz}, or as an affine Lie algebra. The
two-dimensional field theoretic realizations of affine Lie
algebras have been the basis of considerable work in rational
conformal field theory and string theory \cite{go}. While affine
Lie algebras are the best-studied examples of the Kac-Moody
algebras, more generally, they are characterized as follows
\cite{kac,waki}. We can write the generators of any Kac-Moody
algebra, ${\cal G}$, in what is known as the Chevalley basis: the
positive and negative simple root generators, $E_a^{\pm}$, are the
generalization of the raising and lowering operators, $J_{\pm}$,
of the angular momentum group $SU(2)$, familiar to every quantum
physicist. Likewise, the role of the single eigenoperator of
$SU(2)$, usually denoted $J_3$, is more generally played by the
Cartan subalgebra of ${\cal G}$. This is the maximal subset of
mutually-commuting generators, denoted by $H_a$. The number of
generators in the Cartan subalgebra gives the rank, $r$, of the
algebra. We have the familiar commutation relations:
\begin{equation}
[H_a , E_b^{\pm} ] = \pm A_{ab} E_b^{\pm} , \quad [E_{a}^{+} ,
E_b^{-} ] = \delta_{ab} H_a \quad . \label{eq:chev}
\end{equation}
The matrix $A_{ab}$ is known as the Cartan matrix. For the simple
Lie algebras in Cartan's classification, its determinant is
positive-definite: ${\rm det} (A) $ $>$ $0$.  It is important to
note that the Cartan matrix uniquely defines a corresponding
Kac-Moody algebra: given the entries of $A_{ab}$, we can use the
commutation relations in Eq.\ (\ref{eq:chev}) and what are known
as Serre relations:
\begin{equation}
[E_a^+ , \cdots [E^+_a,E^+_b] \cdots ] = 0 , \quad [E_a^- , \cdots
[E^-_a,E^-_b] \cdots ] = 0 \quad , \label{eq:serre}
\end{equation}
to reconstruct the generators and roots of the Kac-Moody algebra.
Labelling the simple roots $\alpha_a$, $a$$=$$1$, $\cdots$, $r$,
where $r$ is the rank of ${\cal G}$, the Cartan matrix can be
parametrized as follows:
\begin{equation}
A_{ab} = 2 {{(\alpha_a , \alpha_b ) }\over{ (\alpha_a , \alpha_a )
}} \quad , \label{eq:bas}
\end{equation}
from which it follows that $A_{aa}$$=$$2$. The generators can be
normalized such that all off-diagonal entries of the Cartan matrix
are negative integers, or zero. The entries of the Cartan matrix
can be conveniently encoded by an unoriented graph with $r$ nodes
and adjacency matrix $2\delta_{ab} - A_{ab}$, known as the Dynkin
diagram. It specifies the Cartan matrix uniquely upto a
re-labelling of rows and columns. Thus, the simply-laced Dynkin
diagrams contain only single links between nodes, since all
off-diagonal entries of $A_{ab}$ are either $-1$, or $0$.
Non-simply-laced Dynkin diagrams can include multiple links,
corresponding to the appearance of other negative integers. A
disconnected Dynkin diagram implies that the Cartan matrix takes
block-diagonal form, and that the algebra decomposes into simple
commuting factors. Notice that the Kac-Moody algebra is invariant
under the set of involutions defined by:
\begin{equation}
E^+_a \to \eta_{a} E_a^- , \quad E^-_a \to \eta_a E^+_a , \quad
H_a \to - H_a \quad , \label{eq:invol}
\end{equation}
where the $\eta_a$$=$$\pm 1$ for every $a$. We emphasize that each
self-consistent choice of phases, $\eta_a$, corresponds to a
distinct Kac-Moody algebra. This freedom in the choice of phases
becomes especially significant when attributing a spacetime
interpretation to the related global symmetry algebra
\cite{west2,keurentjes}, as discussed in Appendix B for $E_{11}$.

\vskip 0.1in Every simple root generator is isomorphic to a vector
$\alpha_a^i$, such that $H_a$$=$$2 {{\alpha_a^i
H_i}\over{(\alpha_a, \alpha_a)}}$, where the $H_i$ define an
alternative basis for the Cartan subalgebra known as the
Cartan-Weyl basis. Thus, to any root, $\alpha$, we can associate
generators, $E_{\alpha}^{\pm}$, such that $[H_i , E_{\alpha}^{\pm}
] = \pm \alpha^i E_{\alpha}^{\pm} $. Notice that each set of three
generators, $\{E_\alpha^+ , \alpha^i H_i , E_\alpha^- \}$, defines
a distinct $A_1$ subalgebra of ${\cal G}$. The $r$ simple root
vectors, $\alpha_a^i$, span an $r$-dimensional vector space known
as the root-lattice of ${\cal G}$, denoted $\Lambda_{\cal G}$. It
is convenient to single out the so-called highest root vector in
this lattice, usually denoted $\theta$, normalized as
$\theta^2$$=$$2$, and parameterized as follows:
$\theta$$=$$\sum_{i=1}^r n_i\alpha_i$, where the $n_i$$=$$2
{{(\theta,\lambda_i)}\over{(\alpha_i,\alpha_i)}}$ are known as the
Kac labels. The $\lambda_i$ are the $r$ distinct fundamental
weight representations of the rank $r$ simple Lie algebra, defined
by the relations $2 {{(\lambda_j , \alpha_i )}\over{(\alpha_i ,
\alpha_j )}}$$=$$\delta_{ij}$ \cite{gilmore,kac,waki}.

\vskip 0.1in Given any finite dimensional semi-simple Lie algebra
${\cal G}$, we can construct its affine extension, ${\cal
G}^{(1)}$, by the addition of a node to its Dynkin diagram. This
construction is reviewed in the classic paper of Goddard and Olive
\cite{go}.\footnote{We have simplified the notation in \cite{go}
\cite{w2} as follows: ${\cal G}^{(n)}$ with $n$$=$$1$, $2$, and
$3$, will denote, respectively, the extension, over-extension, and
very-extension, of the Lie algebra ${\cal G}$. This corresponds to
successive extensions of the rank $r$ simple root basis
$\{\alpha_i , i=1, \cdots , r \}$, by the addition of simple roots
denoted $\alpha_{-n}$, where $n$$=$$1$, $2$, and $3$.} We begin
with the unique Lorentzian even self-dual lattice of dimension
two, $\Pi^{(1,1)}$, with norm $x\cdot y = x_1y_1 - x_{-1} y_{-1}$.
$\Pi^{(1,1)}$ is conveniently expressed in terms of a light-cone
basis, mapping lattice vectors $x,y$ $\to$ $z,w$, where
\cite{go,w2}:
\begin{equation}
z = {{1}\over{{\sqrt{2}} }} (x_1 +  x_{-1}) , \quad {\bar{z}} =
{{1}\over{ {\sqrt{2}} }} (x_1 - x_{-1}), \quad {\rm and} ~ z ,
{\bar z}, w , {\bar {w}} \in (m,n) , \quad \forall ~ m,n \in {\rm
Z} \quad . \label{eq:loret}
\end{equation}
The primitive null vectors of $\Pi^{(1,1)}$ are $k$$=$$(1,0)$ and
${\bar{k}}$$=$$(0,1)$. Let us append $\Pi^{(1,1)}$ to the root
lattice of ${\cal G}$, and consider the subspace of vectors in
$\Lambda_{\cal G} \oplus \Pi^{(1,1)}$ that are orthogonal to the
primitive null vector $k$. It is clear that this subspace includes
all of the root vectors of ${\cal G}$, spanned by the original set
of simple roots $\{ \alpha_i \}$. The enlarged span of the affine
extended algebra is defined by appending an additional simple
root, $\alpha_{-1}$, also referred to as the extended, or affine,
root. Thus, the affine extension of ${\cal G}$ has rank $r+1$. The
extended root takes the form:
\begin{equation}
\alpha_{-1} = k - \theta , \quad \quad k \cdot k =0 , \quad\quad
(\alpha_{-1}, \alpha_{-1}) = 2 \quad  . \label{eq:ext}
\end{equation}
Notice that $\alpha_{-1}$ has been written as an
$(r+2)$-dimensional vector, reflecting the fact that the
root-lattice of ${\cal G}^{(1)}$ is an $(r+1)$-dimensional
projection from the $(r+2)$-dimensional lattice, $\Lambda_{\cal G}
\oplus \Pi^{(1,1)}$. Clearly, the Cartan matrix of the affine
extension of ${\cal G}$ will have one additional row, and one
additional column, with entries given by the scalar products of
$\alpha_{-1}$ with the simple roots of ${\cal G}$:
\begin{equation}
 A_{i,r+1} = 2 {{(\alpha_{-1},\alpha_i)}\over{(\alpha_i ,
\alpha_i )}}, \quad  \quad A_{r+1,i} = (\alpha_{-1},\alpha_i)
\quad . \label{eq:carts}
\end{equation}
It is clear that the determinant of the extended Cartan matrix
vanishes: ${\rm det} A_{ab} ({\cal G}^{(1)} )
$$=$$0$. As an aside, we comment that ${\cal G}^{(1)} $ has
been denoted interchangeably by ${\cal G}^{+}$ in the recent
literature \cite{go,w2,w4,w5}.

\vskip 0.1in This naturally suggests that we ask what rank $(r+2)$
algebra might correspond to the full extension of the
root-lattice, $\Lambda_{\cal G} \oplus \Pi^{(1,1)}$? The
mathematically well-defined way to address this question is to
first return to the Dynkin diagram of ${\cal G}^{(1)}$, adding a
single link to the affine node \cite{go}. This defines what is
known as the over-extension of ${\cal G}$: ${\cal
G}^{(2)}$$=$${\cal G}^{++}$ \cite{go,w2}. The over-extended root
takes the form:
\begin{equation}
\alpha_{-2} = -( k + {\bar{k}} ) , \quad \quad k \cdot k =
{\bar{k}} \cdot {\bar{k}} = 0 , \quad\quad (\alpha_{-2},
\alpha_{-2}) = 2 \quad . \label{eq:oext}
\end{equation}
The Cartan matrix of ${\cal G}^{(2)}$ is extended by the new
entries:
\begin{equation}
 A_{i,r+2} = 0 ,
A_{r+2,i} = 0 ,
 A_{r+1,r+2} = A_{r+2, r+1} = -1 . \quad
\label{eq:carto}
\end{equation}
Notice that the determinant of the Cartan matrix is non-singular
and negative-definite, since ${\cal G}$ was assumed to be a finite
dimensional semi-simple Lie algebra:
\begin{equation}
 {\rm det} A_{{\cal G}^{(2)}} = 2 ~ {\rm det} A_{ {\cal G}^{(1)} } -
 {\rm det} A_{\cal G} = - {\rm det} A_{\cal G} \quad .
\label{eq:ocarts}
\end{equation}
Such a Kac-Moody algebra is said to be {\em Lorentzian}. The
over-extension of a finite dimensional semi-simple Lie algebra is
an especially simple example of a Lorentzian Kac-Moody algebra. We
reiterate that, by construction, the root lattice of ${\cal
G}^{(2)}$ has Lorentzian signature: $\Lambda_{{\cal G}^{(2)}}$ $=$
$\Lambda_{\cal G} \oplus \Pi^{(1,1)}$. The seemingly innocuous
extension to a root-lattice with indefinite norm has profound
consequences. Notice that the root system of a Lorentzian
Kac-Moody algebra includes both real and {\em imaginary} roots,
namely, those with $\beta^2$ $<$ $0$ \cite{kac,nici,ksw,ganor}.
The representation theory of Lorentzian algebras turns out to be
full of surprises: unlike what happens in a finite-dimensional Lie
algebras, the adjoint representation is no longer a highest weight
representation, nor can it be constructed as the tensor product of
fundamentals. In fact, the adjoint representation can contain
within it some of the fundamental representations of the algebra!
We caution the reader that while the representation theory of the
finite-dimensional and affine Lie algebras is known in explicit
detail, only partial results are available in the Lorentzian
cases. But it is encouraging that the standard tools of Kac-Moody
algebras: namely, the characterization of the root system with
respect to the Weyl group of reflections, the use of the Weyl-Kac
character formula for the computation of root multiplicities, and
the well-known Peterson and Freudenthal identities, hold just as
well for the Lorentzian cases \cite{kac,waki}. The explicit
details of the representation theory of the over and very extended
algebras are currently under investigation by a number of groups
\cite{nici,w2,ksw,ganor}.

\vskip 0.1in A further extension of the algebra is enabled by the
addition of a link to the over-extended node. This defines what is
known as the very-extension of ${\cal G}$: ${\cal G}^{(3)}$$=$
${\cal G}^{+++}$, introduced in the work of Olive, Gaberdiel, and
West \cite{w2}. The root-system of the very-extension is given by
the projected subspace of vectors $x$ that are orthogonal to the
primitive timelike vector belonging to an additional $\Pi^{(1,1)}$
factor:
\begin{equation}
x \in \Lambda_{\cal G} \oplus \Pi^{(1,1)} \oplus \Pi^{\prime (1,1)
} , \quad \quad x \cdot ( l - {\bar{l}} ) = 0 , \quad\quad l ,
{\bar{l}} \in \Pi^{\prime (1,1) } \quad , \label{eq:vextpr}
\end{equation}
where $l$, ${\bar{l}}$ are the primitive null vectors of
$\Pi^{\prime (1,1)}$. This root-system is defined by the addition
of the so-called very-extended simple root:
\begin{equation}
\alpha_{-3} = k -( l + {\bar{l}} ) ,  \quad\quad (\alpha_{-3},
\alpha_{-3}) = 2 , ~ (\alpha_{-3},\alpha_{-2})= -1,~
(\alpha_{-3},\alpha_{-1})=(\alpha_{-3},\alpha_i) = 0 \quad .
\label{eq:vext}
\end{equation}
Recall that $k$ is a primitive null vector of $\Pi^{(1,1)}$. It is
easy to see that the Cartan matrix of the very extension simply
corresponds to the addition of a row, and column, with mostly
zeroes, plus the single nonvanishing off-diagonal entries,
$A_{r+2,r+3}$$=$$A_{r+3,r+2}$$=$$-1$. The determinant of the
Cartan matrix is, once again, negative-definite: ${\rm det}
A_{{\cal G}^{(3)}}$$=$$- 2 {\rm det} A_{\cal G}$.

\vskip 0.1in The weight system of the very-extended algebra can be
inferred by tracing its progression thru the iterative
construction described above \cite{w2}. The result is easy to
motivate. In terms of the fundamental weights of the finite
dimensional Lie algebra, $\lambda_i$, $i$$=$$1$, $\cdots$, $r$, we
have:
\begin{eqnarray}
\lambda_i =&& \lambda_i - (\lambda_i , \theta) [ k - {\bar{k}} -
{{1}\over{2}} (l + {\bar{l}} ) ] \cr \lambda_{-1} =&& - [ k -
{\bar{k}} - {{1}\over{2}} (l + {\bar{l}} ) ] , \quad \lambda_{-2}
= - k , \quad \lambda_{-3} = - {{1}\over{2}} (l + {\bar{l}} )
\quad . \label{eq:wgts}
\end{eqnarray}
The weights of the over-extended algebra can be recovered from
these expressions by simply setting $l$$=$${\bar{l}}$$=$$0$. For a
simply-laced finite-dimensional Lie algebra with simply-laced
root-lattice, and dual weight-lattice \cite{gilmore}, it is easy
to write down the weight-lattice of the over- and very- extensions
\cite{w2}. The weight-lattice of the over-extension ${\cal
G}^{(2)}$ is given by:
\begin{equation}
[\Lambda_{{\cal G}^{(2)}}]^* = \Lambda_{\cal G}^* \oplus
\Pi^{(1,1)}, \quad \quad\quad \rightarrow ~ {{[\Lambda_{{\cal
G}^{(2)}}]^*}\over{\Lambda_{{\cal G}^{(2)}} }} = {\rm Z}_{\cal G}
\quad . \label{eq:wej}
\end{equation}
Likewise, for the very-extended algebra, ${\cal G}^{(3)}$, the
root-lattice and weight-lattice, respectively, take the form:
\begin{equation}
\Lambda_{{\cal G}^{(3)}} = \Lambda_{\cal G} \oplus \Pi^{(1,1)}
\oplus \{ (m,-m) : m \in {\rm Z} \} , \quad [\Lambda_{{\cal
G}^{(3)}}]^* = \Lambda_{\cal G}^* \oplus \Pi^{(1,1)} \oplus \{
(n,-n) : 2n \in {\rm Z} \}  \quad . \label{eq:wjkl}
\end{equation}
Notice that the roots and weights of the rank $(r+3)$ algebra are
expressed here as vectors in an $(r+4)$-dimensional vector space.
We can infer that their coset takes the form:
\begin{equation}
{{[\Lambda_{{\cal G}^{(3)}}]^*}\over{\Lambda_{{\cal G}^{(3)}} }} =
{\rm Z}_{\cal G} \times {\rm Z}_2 \quad . \label{eq:wejls}
\end{equation}
A more pedestrian approach to the representation theory of very
extended algebras that eschews the traditional, and more abstract,
methodology of the Weyl-Kac character formula and Freudenthal
identity, can be found in \cite{w2,ksw,kw,w6}. It has become
conventional to label the Dynkin diagram of the very extended
algebra as follows: the very, over, and affine, nodes are labelled
1, 2, and 3, starting from left to right along the horizontal, and
then continuing from right to left with any nodes above the
horizontal. Notice that deletion of a single node of the Dynkin
diagram of the very extended algebra, also called the central
node, always gives the Dynkin diagram of a finite-dimensional Lie
algebra. In the case of $E_8^{(3)}$, the central node is labelled
11, and its deletion gives the Dynkin diagram of its $A_{10}$
gravity subalgebra. Thus, any generic root, $\beta$, has a simple
root decomposition that takes the form:
\begin{equation}
\beta = n_c \alpha_c + \sum_{i} n_i \alpha_i \quad ,
\label{eq:centr} \end{equation} where the $\alpha_i$ are the
simple roots of the finite-dimensional algebra following deletion
of the central node. The integer $n_c$ is defined as the {\em
level} of the Lorentzian Kac-Moody algebra
\cite{w2,nici,ksw,kw,w6}. Roots at level zero are simply those of
the corresponding finite-dimensional Lie subalgebra. It should be
noted that the commutators of the algebra preserve the level.

 \vskip 0.1in\noindent{\bf Very-Extension of $E_8$}: The cases of
interest in this paper are the affine-, over-, and very-
extensions of the simply-laced Lie algebra $E_8$, with its famous
rank eight Euclidean even self-dual root-lattice
\cite{gilmore,polbook}. In arriving at the Dynkin diagram of
$E_{11}$ $=$ $E_8^{(3)}$, we first construct the affine extension
of $E_8$, which is known as $E_9$, followed by its over-extension,
$E_{10}$, of rank ten. This is the highest rank example of the
hyperbolic Kac-Moody algebras. They have been exhaustively
classified in the mathematics literature \cite{julia,ganor}.
$\Lambda_{E_8}$ is spanned by the following eight simple root
vectors \cite{gilmore}:
\begin{eqnarray}
\alpha_1 =&& (1,+1,0,0,0,0,0,0) \quad \alpha_2 =
(1,-1,0,0,0,0,0,0) \cr \alpha_3 =&& (0,1,-1,0,0,0,0,0) \quad
\alpha_4 = (0,0,1,-1,0,0,0,0) \cr \alpha_5 =&& (0,0,0,1,-1,0,0,0)
\quad \alpha_6 = (0,0,0,0,1,-1,0,0) \cr \alpha_7 =&&
(0,0,0,0,0,1,-1,0) \quad \alpha_8 =
(\half,\half,\half,\half,\half,\half,\half,\half) \quad .
\label{eq:latts}
\end{eqnarray}
Appending the extended root $\alpha_{-1}$ determines the affine
extension, $E_9$, where we substitute for the highest root,
$\theta$, with the result:
\begin{equation}
\alpha_{-1} = k - \theta = ((0,0,0,0,0,0,-1,1),(1,0)) \quad .
\label{eq:e9}
\end{equation}
The over-extension, $E_{10}$, is specified by including the
additional simple root:
\begin{equation}
\alpha_{-2} = -( k + {\bar{k}}) = ((0,0,0,0,0,0,0,0),(-1,1)) \quad
. \label{eq:e10}
\end{equation}
The rank-ten root-lattice of $E_{10}$, namely, $\Lambda_{E_8}
\oplus \Pi^{(1,1)}$, is even, and self-dual, by construction. It
coincides with the unique even Lorentzian self-dual lattice of
dimension ten, usually denoted $\Pi^{(9,1)}$. Finally, for the
very-extension, $E_{11}$$=$$E_8^{(3)}$, we include the additional
simple root:
\begin{equation}
\alpha_{-3} = k -( l + {\bar{l}}) =
((0,0,0,0,0,0,0,0),(1,0),(-1,1)) \quad . \label{eq:e11}
\end{equation}
The root system of $E_{11}$ is the projected subspace orthogonal
to the primitive timelike vector, $l+ {\bar{l}}$, in the unique
even Lorentzian self-dual lattice in dimensions $(10,2)$:
\begin{equation}
x \in \Pi^{(10,2)} = \Lambda_{E_8} \oplus \Pi^{(1,1)} \oplus
\Pi^{(1,1)} , \quad \quad x \cdot (l - {\bar{l}} ) = 0 \quad .
\label{eq:e11s}
\end{equation}
As a consequence of the self-duality property of the $E_8$
lattice, we now have the elegant result:
\begin{equation}
{{[\Lambda_{E_{11}}]^*}\over{\Lambda_{E_{11} } }} = {\rm Z}_2
\quad . \label{eq:wejlse11}
\end{equation}
In summary, notice that, by construction, the rank-eleven algebra
$E_{11}$ contains the full Cremmer-Julia $E_{11-n}$, $n$$=0$,
$\cdots$, $11$, sequence of hidden symmetry algebras as proper
subalgebras. This was, in fact, the original motivation for West's
construction \cite{west2}. But it is a beautiful consequence that
$E_8^{(3)}$ also incorporates the crucial nine-form potential: the
generator associated with the very-extended node of its Dynkin
diagram, thus unifying Roman's massive type IIA supergravity with
both M theory, as well as the massless IIA and IIB supergravities
and their toroidal compactifications to lower spacetime
dimensions. This is precisely as was required by our conventional
understanding of string/M dualities
\cite{sen,hulltown,witten,dbrane,polbook}.

\vskip 0.4in \noindent{\large{\bf Appendix B: The Nonlinear
Realization of the $E_{11}$ $=$ $E_8^{(3)}$ Algebra}}

\vskip 0.1in The review paper by Cremmer, Julia, Lu, and Pope
\cite{cjlp} contains a detailed explanation of the method of
nonlinear realizations from first principles in section 4, and we
urge the non-specialist to consult this reference prior to reading
the recent work on very-extensions
\cite{west1,west2,w0,w1,w2,w4,w5,kw}. For completeness, we begin
with a brief overview of the straightforward nonlinear realization
of the scalar Lagrangian in dimensions $D$ $\ge$ $6$ in this
appendix. The more complicated analysis for the cases $3$ $\le$
$D$ $\le$ $5$ can be found in the references \cite{cj,cjlp,lp}.

\vskip 0.1in We begin with the Lagrangian of the fully-undualized
toroidally-compactified eleven-dimensional supergravity in $D$
dimensions, using the notation in \cite{cjlp}. Let us denote the
set of dilaton vectors as ${\cal A}^i_{[0]j}$ $=$ $b_{ij}$, and
the axions collectively as ${\cal A}_{[0]ijk}$ $=$ $a_{ijk}$. The
$i,j,k$ are internal indices taking values from $1$, $\cdots$,
$11$$-$$D$; at this stage, one need not distinguish them from
tangent space indices. The key observation, dating back to
\cite{cj}, is that in each case the scalars are in one-to-one
correspondence with the positive root vectors of the $E_{11-n}$
algebra:
\begin{eqnarray}
{\cal L} =&& eR - {{1}\over{2}} e (\partial {\bf \phi})^2 -
{{1}\over{48}} e e^{\bf a \cdot \phi} F_{[4]}^2 - {{1}\over{12}} e
\sum_i e^{{\bf a_i \cdot \phi}} (F_{[3]i})^2 - {{1}\over{4}} e
\sum_{i<j} e^{{\bf a_{ij} \cdot \phi}} (F_{[2]ij})^2 \cr \quad &&-
{{1}\over{4}} e \sum_i e^{{\bf b_i \cdot \phi}} (F^i_{[2]})^2  -
{{1}\over{2}} \sum_{i<j<k} e^{{\bf a_{ijk} \cdot \phi}}
(F_{[1]ijk})^2 - {{1}\over{2}} e \sum_{i<j} e^{{\bf b_{ij}\cdot
\phi}} ({\cal F}^i_{[1]j})^2 + {\cal L}_{\bf C-S} \quad ,
\label{eq:lagra}
\end{eqnarray}
where ${\cal L}_{\rm CS}$ is the dimensional reduction of the
$F_{[4]}$$\wedge$$F_{[4]}$$\wedge$$A_{[3]}$ in eleven dimensions.
The notation distinguishes the 1-form field strengths by their
origin in the metric, ${\cal F}^i_{[1]j}$, or in the three-form
potential, $ F_{[1]ijk}$, of eleven-dimensional supergravity:
\begin{equation}
F_{[1]ijk} = d A_{[0]ijk} - {\cal A} \wedge d {\cal A} ~ {\rm
terms}, \quad {\cal F}^i_{[1]j} = d {\cal A}^i_{[0]j} - {\cal A}
\wedge d {\cal A} ~ {\rm terms} \quad . \label{eq:forms}
\end{equation}
We choose $b_{i,i+1}$ and $a_{123}$ as simple roots; removing
$a_{123}$ gives the simple roots of $SL(11-D,{\bf R})$. As is
usual, the root-lattice is generated by linear combinations of the
simple roots with positive-definite integer coefficients. The
Dynkin diagram is as shown in Figure 1. Note that inclusion of the
axions from dualized field-strengths in $3$ $\le$ $D$ $\le$ $5$,
namely, $({\bf a},{\bf a}_i, {\bf b}_i, {\bf a}_{ij})$, fill out
the root-lattices of $E_6$, $E_7$, and $E_8$. It is helpful to
introduce the following parameterization of roots \cite{lp,cjlp}:
\begin{equation}
{\bf a}=-{\bf g}, \quad {\bf b}_i = -{\bf f}_i, \quad  {\bf a}_i =
{\bf f}_i - {\bf g}, \quad {\bf b}_{ij}={\bf f}_j -{\bf f}_i ,
\quad {\bf a}_{ij}={\bf f}_i + {\bf f}_j - {\bf g}, \quad {\bf
a}_{ijk} = {\bf f}_i + {\bf f}_j + {\bf f}_k - {\bf g} \quad ,
\label{eq:para}
\end{equation}
where ${\bf f}_i$ and ${\bf g}$ are $(11-D)$-dimensional vectors
satisfying the relations: ${\bf g} \cdot {\bf g} =
2{{11-D}\over{D-2}}$, ${\bf g} \cdot {\bf f}_i = {{6}\over{D-2}}$,
and ${\bf f}_i \cdot {\bf f}_j = 2 \delta_{ij} + {{2}\over{D-2}}$.
Also, $\sum {\bf f}_i$$=$$3{\bf g}$.  It follows that ${\bf
b}_{ik}$$=$${\bf b}_{ij}$$+$${\bf b}_{jk}$, and ${\bf
a}_{ijk}$$+$${\bf b}_{il}$$=$${\bf a}_{ljk}$. Identifying the
positive roots ${\bf b}_{ij}$ and ${\bf a}_{ijk}$ with generators
$E^i_j$ and $E^{ijk}$, respectively, we see that they satisfy the
commutation relations:
\begin{equation}
[E_i^j , E_k^l ] = \delta^j_k E_i^l - \delta_i^l E_k^j , \quad
[E_l^m , E^{ijk} ] = -3 \delta_l^{[i} E^{|m|jk]} , \quad [E^{ijk}
, E^{lmn} ] = 0 \quad . \label{eq:crn}
\end{equation}
The first two relations are an expression of $SL(11-D,{\bf R})$
covariance. In dimensions $D$$\ge$$6$, the generators $E^{ijk}$
commute. For $D$$\le$$5$, whether or not they commute depends upon
dualizations. Finally, if we include the hidden subgroup ${\bf
R}_s$, writing the Cartan generators as a vector ${\bf H}$, we
have:
\begin{equation}
[{\bf H} , E_i^j ] = {\bf b}_{ij} E_i^j , \quad [{\bf H} , E^{ijk}
] = {\bf a}_{ijk} E^{ijk} \quad {\rm no ~ sum} \quad .
\label{eq:cartsg}
\end{equation}
Introducing the non-linear realization \cite{cj,cjlp}:
\begin{equation}
{\cal V} = e^{ {{1}\over{2}}{\bf \phi}\cdot{\bf H} } e^{{\bf
b}_{ij} E_i^j } e^{\sum_{i<j<k} {\bf a}_{ijk} E^{ijk}} \quad ,
\label{eq:mcf}
\end{equation}
it can be verified that:
\begin{equation}
d {\cal V} {\cal V}^{-1} = {{1}\over{2}} d{\bf \phi } \cdot {\bf
H} + \sum_{i<j} e^{ {{1}\over{2}} {\bf b}_{ij} \cdot {\bf \phi} }
{\cal F}^i_{[1]j} E_i^j + \sum_{i<j<k} e^{ {{1}\over{2}} {\bf
a}_{ijk} \cdot {\bf \phi} } F_{[1]ijk} E^{ijk} \quad ,
\label{eq:mcfl}
\end{equation}
and the entire scalar Lagrangian is expressible in the form:
\begin{equation}
{\cal L} = {{1}\over{4}} tr \left ( \partial {\cal M}^{-1}
\partial {\cal M} \right ) \quad ,
\label{eq:lagb}
\end{equation}
where we define ${\cal M} = {\cal V}^T {\cal V}$, and where the
superscript denotes the transpose. Having written the scalar
Lagrangian in Meurer-Cartan form, it is helpful to identify the
remnant local gauge symmetry. The transformation:
\begin{equation}
{\cal M} \to {\cal M}^{\prime} = U^T {\cal M} U \quad ,
\label{eq:symm}
\end{equation}
where $U$ is a constant element in the global symmetry group, is
found to leave the Lagrangian invariant. Thus, the Lagrangian is
made out of $K(E_{11-D})$ invariants, where $K(G)$ is the maximal
compact subgroup of $G$, and we have the coset structure $G/K(G)$
for $D$ $ \ge$ $6$.

\vskip 0.1in West's nonlinear realizations of the hidden symmetry
algebras underlying the different supergravity theories and M
theory is similar in spirit, but brings in many new features
\cite{west1,west2,w4}. Earlier in the text, a Hodge dual was
introduced for each $p$-form gauge potential, other than the
self-dual potentials, in addition to the generators of $GL(n,{\rm
R})$, the translations, $P^a$, and scalars, $R_{s}$, if present in
the supergravity theory. In the case of M theory, with its
eleven-dimensional supergravity field theoretic limit, West
proposes the following realization of an $E_{11}$ $=$ $E_8^{(3)}$
algebra:
\begin{equation}
[K^a_b , K^c_d ] = \delta_b^c K^a_d - \delta_d^a K^c_b ,
\quad\quad [K^a_b , R^{c_1 \cdots c_p } ] = \delta_b^{c_1} R^{ac_2
\cdots c_p} + \cdots , \quad \quad p=3, 6 \quad , \label{eq:rels}
\end{equation}
along with the additional commutator:
\begin{equation}
 [K^a_b ,
R^{c_1 \cdots c_8,d } ] = \left ( \delta_b^{c_1} R^{ac_2 \cdots
c_8,d} + \cdots \right ) +
 \delta_b^d R^{c_1 \cdots c_8,b} \quad , \label{eq:gl1}
\end{equation}
The $K^a_b$ are the generators of the $A_{10}$ subalgebra of
$E_8^{(3)}$. Notice that this only represents the
volume-preserving subgroup, $SL(11,{\bf R})$, of the expected
$GL(11,{\bf R})$, and, in addition, that the momentum generators
of eleven-dimensional supergravity are not part of the
$E_{8}^{(3)}$. Thus, from this algebraic perspective, if $E_{11}$
is indeed the symmetry algebra of M theory, it does not appear to
be an inherently eleven-dimensional theory.\footnote{We are
describing the most recent formulation given in \cite{w4}. We
should warn the reader that there has been a significant shift in
perspective from West's earliest proposal regarding $E_{11}$,
namely, in \cite{west1}, to the more recent ideas summarized in
\cite{w2,w4}. West remarks in \cite{w4} that a better name for
this conjectured high-energy completion of eleven-dimensional
supergravity might be {\em E-theory}.}

\vskip 0.1in The $+\cdots$ in the commutation relations denotes
the antisymmetrization of all indices. Notice that the nine-index
generator is antisymmetrized in only the first 8 indices. There
is, of course, no nine-form potential in eleven-dimensional
supergravity. The inclusion of a nine-index generator in the
purported symmetry algebra is, therefore, a definitive statement
that M theory is more than its low-energy limit. A more physical
motivation is Roman's IIA mass parameter: inclusion of the
nine-index generator in the symmetry algebra of M theory enables a
simple relationship with the hidden symmetry algebra of the
massive type IIA supergravity described in the introduction. In
addition, note that $E_{11}$ contains as proper subalgebras the
entire Cremmer-Julia $E_{11-n}$ sequence. As an aside, the
subalgebra generated by $R^{c_1c_2c_3}$ and $R^{c_1 \cdots c_6}$
alone was previously identified as the global symmetry algebra of
the M5brane in \cite{wfive}, an early motivation for this
algebraic approach to M theory. Thus, we have:
\begin{equation}
 [R^{c_1 \cdots c_3} , R^{c_4 \cdots c_6} ] = 2  R^{c_1
\cdots c_{6}},
 \quad  [R^{c_1 \cdots c_6} , R^{b_1 \cdots b_3} ] = 3R^{c_1 \cdots c_6[b_1 b_2,b_3]},
 \quad  [R^{c_1 \cdots c_6} , R^{b_1 \cdots b_6} ] = 0 \quad .
\label{eq:extrar}
\end{equation}
The Chevalley generators corresponding to positive simple roots of
$E_{8}^{(3)}$ can be identified as follows \cite{w4}:
\begin{equation}
E_a = K^a_{a+1} , \quad a=1, \cdots , 10 , \quad E_{11} = R^{9 10
11} \quad , \label{eq:checv}
\end{equation}
and the rank eleven Cartan subalgebra is generated by:
\begin{equation}
H_a = K^a_a - K^{a+1}_{a+1} , ~ a = 1, \cdots , 10 , \quad H_{11}
= - {{1}\over{3}} (K^1_1 + \cdots + K^8_8 ) + {{2}\over{3}} (K^9_9
 + K^{10}_{10} + K^{11}_{11} ) \quad .
\label{eq:cart}
\end{equation}
Notice that the six-form and nine-index generator do not enter at
this level (zero) of the Kac-Moody algebra. However, since the
commutator of $R^{91011}$ with generic $K^a_b$ generates all of
the remaining components of the three-form potential, the
commutator of the three-form with itself generates all components
of the six-form and, finally, the commutators of the six-form and
three-form yield the components of the nine-index generator, we
can span the root-system of $E_8^{(3)}$ with this simple choice of
basis. Thus, any generic root in the root-system will be
isomorphic to a string of commutators of Chevalley generators,
mapping to a unique group element via the nonlinear realization to
which we now turn. The non-linear realization of $E_{8}^{(3)}$ is
built up from group elements that take the form:
\begin{equation}
g = \exp \left [ \sum_{a \le b} h^a_b K^b_a \right ] \exp \left [
\thirdf A_{c_1c_2c_3} R^{c_1 c_2c_3} \right ]  \exp \left [
\sixthf A_{c_1 \cdots c_6} R^{c_1 \cdots c_6} \right ] \exp \left
[ \eigthf h_{c_1 \cdots c_8,d} R^{c_1 \cdots c_8,d}  \right ]
\quad , \label{eq:nlmg}
\end{equation}
$A_{[3]}$, and $A_{[6]}$, are to be identified, respectively, with
the three-form potential of supergravity, and its Hodge dual.
$h^a_b$ is related to the vielbein as shown below, and $h_{c_1
\cdots c_8 , d}$ plays the role of a dual field of gravity. Thus,
unlike previous proposals such as the doubled-field formalism
\cite{cjlp}, the notion of duality, and the framework of
non-linear realizations, has been extended to the {\em full}
bosonic sector, inclusive of the graviton! West's motivation for a
dual-field formalism for gravity comes from an older work by
Borisov and Ogievetsky \cite{bo}, and is described in
\cite{west1,west2}. Notice that, since the generators of spacetime
translations were absent from the $E_8^{(3)}$ algebra, they also
do not appear in the group element. This is unlike the proposed
nonlinear realizations of the ten-dimensional type II
supergravities written down in \cite{west1,w0,w1}, which are based
on the ${\cal G}_{\rm II}$ algebras described in the introduction,
and which explicitly include translations. Although it is possible
to invoke the nonlinear realization of $E_8^{(3)}$ to develop an
isomorphism of group elements to specific eleven-dimensional line
elements because of the appearance of the vielbein, as will be
illustrated below, the emergence of the translation generators
from this algebraic framework remains an interesting puzzle
\cite{w4,kw}. It is discussed further in section 4.

\vskip 0.1in Let us now discuss the significance of the choice of
phases $\eta_a$$=$$\pm 1$ in Eq.\ (32), without which the bosonic
$E_8^{(3)}$ algebra has not been unambiguously specified. As shown
by West, taking $\eta_1$$=+1$, and all remaining $\eta_a$ negative
\cite{west2}, leads to a hidden symmetry algebra with both an
appropriate supersymmetric extension and a spacetime with
Minkowskian signature and a single time direction, $(-,+,\cdots ,
+)$. Taking all of the $\eta_a$ negative gives, instead, a
spacetime of Euclidean signature, defining what is known as the
Cartan involution-invariant subalgebra \cite{west2}. This is also
a physically well-motivated choice, of obvious relevance to future
discussions of string/M theory at finite temperature \cite{cosmo}.
Note that West interchangeably invokes either choice of phase in
later work \cite{w4}, since it is clear that the two alternatives
simply correspond to a Wick rotation in spacetime.

\vskip 0.1in The fact that every other choice of phases in the
bosonic $E_{11}$ algebra leads to an indefinite spacetime metric
with two, or more, timelike directions, some of which do not even
admit supersymmetric extension, was clarified in a recent paper by
Keurentjes \cite{keurentjes}. Restricting to $E_{11}$ algebras
that also admit a supersymmetric extension, one finds new
self-consistent choices of phase correspond to indefinite
spacetimes of signature (2,9), (5,6), (6,5), or (9,2)
\cite{keurentjes}. In other words, the relationship between the
supersymmetric extension of an $E_{11}$ algebra and M theory is
unique upto freedom in the signature of spacetime. Interestingly,
these signatures can be identified with the conjectured $M^*$ and
$M^{\prime}$ eleven-dimensional theories of Hull \cite{hulld},
whose existence was inferred by the application of timelike
dualities on the standard (1,10) signature spacetime metric for
eleven-dimensional supergravity. This clarifies the important fact
that Hull's new solutions \cite{hulld} do not correspond to
distinct eleven-dimensional theories: they belong in a theory with
identical hidden symmetries, apart from the different spacetime
signature.\footnote{Of course, it is not clear at the present time
that any physical significance should be attributed to these
alternative solutions, in which case the corresponding $E_{11}$
algebras can eventually be discarded.} It is a most satisfying
result following from Keurentjes' analysis that invoking either
Euclidean (0,11) or Minkowskian (1,10) spacetime signature--- each
of which has a clear-cut physical interpretation, {\em uniquely}
determines a bosonic $E_{11}$ algebra with unambiguous phase
choice. Note also that while the Minkowskian case permits
supersymmetric extension, the Euclidean algebra does not,
precisely as required by its physical interpretation as the
symmetry algebra of finite temperature M theory.

\vskip 0.1in We will conclude this appendix by illustrating: (i)
the isomorphism between a specific group element, $g$, and the
line-element of a well-known classical background of supergravity.
(ii) the origin of the D8brane of the massive IIA supergravity,
and the space-filling D9brane of the IIB supergravity, in specific
group elements of $E_8^{(3)}$. We urge the reader to consult
\cite{w4} for many more examples of such isomorphisms. The field
$h^a_b$ is related to the vielbein as follows:
\begin{equation}
e^{\mu}_a = e^{h^b_b} (e^{ {\bar{h}} })_{\mu}^a ~ ,
\quad\quad\quad  {\rm where} ~ {\bar{h}}^a_b = h^a_b - \delta^a_b
h^c_c \quad , \label{eq:vielbe}
\end{equation}
We start with our favourite line element, for example, the M2brane
metric discovered by Duff and Stelle \cite{ds}:
\begin{equation}
ds^2 = N_2^{-2/3} (-(dx_1)^2 + (dx_2)^2 +(dx_3)^2 ) + N_2^{1/3}
((dx_4)^2 + \cdots (d x_{11})^2 ) \quad , \label{eq:linem2}
\end{equation}
with four-form field strength, $F_{1234}$ $=$ $\partial_m
N_2^{-1}$, where $m$ labels directions $4$, $\cdots$, $11$,
orthogonal to the worldvolume of the M2brane, and the harmonic
function $N_2$ $=$ $1 + k/r^2$, and $r^2$ $=$ $\sum_{m=4}^{11}
(x_m)^2$. Thus, we can identify:
\begin{equation}
(e^h)^1_1 = (e^h)^2_2 = (e^h)^3_3 = N_2^{-1/3} , \quad (e^h)^4_4 =
\cdots = (e^h)^{11}_{11} = N_2^{-1/6} , \quad A_{123} = N_2^{-1} -
1 \quad , \label{eq:comps}
\end{equation}
where it should be noted that the gauge field is specified with
respect to tangent space. Substituting into Eq.\ (\ref{eq:nlm2}),
we construct the corresponding group element:
\begin{equation}
g_{\rm M2} = \exp \left [ - \half {\rm ln} N_2 \left ( \twothird (
K^1_1 +K^2_2 + K^3_3 ) -  \third (K_4^4 + \cdots K_{11}^{11}
 ) \right ) \right ] \exp \left [ (1 - N_2 ) R^{123} \right ] \quad , \label{eq:nlm2}
\end{equation}
As explained in \cite{w4}, such isomorphisms extend to a vast
spectrum of classical supergravity backgrounds, including bound
states of multiple branes. Based on a large number of examples
\cite{w4}, West argues that the group element corresponding to a
specific half-BPS solution of supergravity parameterized by a
harmonic function $N$ can always be written in the form:
\begin{equation}
g = \exp \left [ - \half {\rm ln} N \beta \cdot H \right ] \exp
\left [ (1-N) E_{\beta} \right ] \quad , \label{eq:line}
\end{equation}
where $\beta$ is the corresponding root, and $E_{\beta}$ the
corresponding generator in $E_8^{(3)}$. This is a most remarkable
identification.

\vskip 0.1in Perhaps even more striking from our perspective is
the detailed correspondence developed in \cite{w4} between the
generators of $E_{8}^{(3)}$, and those of the global symmetry
algebras of the maximal ten-dimensional supergravities, namely,
${\cal G}_{\rm mIIA}$ and ${\cal G}_{\rm IIB}$. West begins by
observing that the Dynkin diagram of $E_8^{(3)}$ has precisely two
inequivalent $A_{9}$ subalgebras. These are distinguished with
respect to the bifurcation at the very-extended node, which can be
labelled 8 on the Dynkin diagram of either $A_{9}$. Decomposing
$E_8^{(3)}$ with respect to the $A_9$ subalgebra corresponding to
the IIA theory, we identify the following Chevalley basis:
\begin{equation}
E_a = K^a_{a+1} , \quad a= 1, ~ \cdots, ~ 9 , \quad E_{10} =
R^{10} , \quad E_{11} = R^{910} \quad , \label{eq:cheviia}
\end{equation}
and corresponding Cartan subalgebra:
\begin{eqnarray}
H_a =&& K^a_a - K^{a+1}_{a+1} , \quad\quad a = 1, \cdots , 9  \cr
H_{10} =&& - \eigth (K^1_1 + \cdots + K^9_9 ) + \eigth K^{10}_{10}
- \threehalf R \cr  H_{11} =&& - \quart (K^1_1 + \cdots + K^8_8) +
\tquart (K^9_9 + K^{10}_{10} ) + R \quad . \label{eq:carti}
\end{eqnarray}
Notice that this only preserves an $SL(10,{\bf R})$ subgroup as
appropriate for a ten-dimensional theory. $R$ is the IIA dilaton,
and $R^{10}$ and $R^{910}$ are, respectively, components of the
IIA R-R one-form and NS-NS two-form generators. By inspection of
the commutation relations among the $K^a_b$ and $p$-form
generators given in Eq.\ (\ref{eq:gl}), and Eq.\ (\ref{eq:extra}),
it is easy to see that this choice of basis suffices to generate
the full set of $p$-form generators entering the mIIA global
symmetry algebra. Thus, if we exclude spacetime translations,
preserving only the $SL(10,{\bf R})$ subalgebra of $GL(10,{\bf
R})$, we find a clear-cut isomorphism between the generators of
$E_8^{(3)}$, and a restricted subset of the generators of ${\cal
G}_{\rm mIIA}$: $\{ (K^a_b,~ a,b=1, ~ \cdots,~ 10), ~ R,~ (R^{c_1
\cdots c_q}, q=1, ~ \cdots , 9) \}$. As argued by us in the main
text, the symmetry algebra of significance to Matrix Theory is
precisely this restriction of ${\cal G}_{\rm mIIA}$.

\vskip 0.1in Clearly, it is therefore possible to identify a
specific $E_8^{(3)}$ group element corresponding to each member in
the spectrum of $(p+1)$-form generators in ${\cal G}_{\rm mIIA}$.
But, most remarkably, upon substitution in Eq.\ (\ref{eq:line}),
West succeeds in deducing the line element for the corresponding
p-brane in full agreement with previous results \cite{w4}. We
begin by noting that the $p$-brane must couple to a $(p+1)$-form
potential with corresponding generator in the $E_8^{(3)}$ algebra.
We begin with the non-linear realization \cite{w4}:
\begin{equation}
g_{\rm mIIA} = \exp \left [ \sum_a h^a_a K^a_a \right ] \exp \left
[ \sum_{a < b} h^a_b K^a_b \right ] \prod_{q=9}^5 \exp \left [
\qfac A_{c_1 \cdots c_q} R^{c_1 \cdots c_q} \right ] \prod_{q=3}^1
\exp \left [ \qfac A_{c_1 \cdots c_q} R^{c_1 \cdots c_q} \right ]
\exp \left [ A R \right ] \quad , \label{eq:griia}
\end{equation}
where special attention must be paid to the reverse ordering in
the products, with the zero-form acting first and the nine-form
acting last. Suppose we wish to deduce the line-element of the
supergravity $p$-brane coupling to the $(p+1)$-form field with
corresponding root, $\beta_{p+1}$, and corresponding lowest weight
generator: $E_{\beta_{p+1}}$ $=$ $R^{1 \cdots p+1 }$. By
inspection of the commutation relations, we can identify this
generator as a string of commutators starting with the positive
and negative simple roots in the Chevalley basis. This identifies
the corresponding $E_8^{(3)}$ root \cite{w4}:
\begin{equation}
\beta_{p+1} \cdot H = \left \{ \eigth (7-p) \left ( K^1_1 + \cdots
+ K_{p+1}^{p+1} \right ) - \eigth (p+1) \left ( K^{p+2}_{p+2} +
\cdots + K^{10}_{10} \right ) \right \} + b_p R \quad ,
\label{eq:cartgdb}
\end{equation}
where $b_7$ $=$ $0$, and $b_p$ $=$ $\half \eta (p-3)$, for $p$
$\le$ $6$, and with $\eta$$=$$\pm 1$ for NS-NS, R-R, respectively.
Substituting in Eq.\ (\ref{eq:line}), we have the result:
\begin{equation}
g_p = \exp \left [ - \half {\rm ln} N \left \{ \cdots \right \} -
\half b_p {\rm ln} N_p R \right ] \exp \left [ (1-N_p )
E_{\beta_{p+1}} \right ] \quad , \label{eq:linep}
\end{equation}
where $\left \{ \cdots \right \}$ represents the linear
combination of generators appearing within curly brackets on the
R.H.S. of Eq.\ (\ref{eq:cartgdb}). Using the identity:
\begin{equation}
\exp \left [ - \half b_p {\rm ln} N_p  \right ] \exp \left [
(1-N_p ) E_{\beta_{p+1}} \right ] = \exp \left [ N_p^{- \half b_p
c_{p+1} } (1-N_p ) E_{\beta_{p+1}} \right ]\exp \left [ - \half
b_p {\rm ln} N_p R \right ] \quad , \label{eq:ident}
\end{equation}
where the $c_{p+1}$$=$$\quart \eta (p-3)$ dependence arises from
the $[R, R^{p+1}]$ commutator. We can read off the solution for
the dilaton and $(p+1)$-form potential in terms of the harmonic
function $N_p$:
\begin{equation}
e^A = (N_p)^{- \half b_p}  , \quad\quad A_{1 \cdots p+1} =
N_p^{-1} - 1 \quad . \label{eq:dilat}
\end{equation}
The corresponding line elements take the form:
\begin{equation}
ds^2 = N_p^{- \eigth (7-p)} (- (dx_1)^2 + (d x_2)^2 + \cdots + (d
x_{p+1})^2 ) + N_p^{\eigth (p+1)} ( (dx_{p+2})^2 + \cdots (d
x_{10})^2 ) \quad . \label{eq:linee}
\end{equation}
Thus, we recover the well-known results for the line elements of
the half BPS pbranes of type IIA supergravity \cite{dl,hs}: here,
p=1, 5 parameterize the fundamental string and NS5brane, while
p=0,2,4,6 parameterize Dpbranes in the R-R sector. The solution
with p=7 is ordinary Minkowskian spacetime.

\vskip 0.1in Moving on to the IIB theory with global symmetry
algebra ${\cal G}_{\rm IIB}$ described in Eqs.\ (\ref{eq:IIB}),
(\ref{eq:extrab}), and (\ref{eq:strb}), we make a corresponding
identification of Chevalley generators with inequivalent choice of
the $A_9$ subalgebra:
\begin{equation}
E_a = K^a_{a+1} , \quad a= 1, ~ \cdots, ~ 8 , \quad E_{11} =
K^9_{10} , \quad E_{9} = R_1^{910} , \quad E_{10} = R_2 \quad ,
\label{eq:cheviib}
\end{equation}
and corresponding Cartan subalgebra:
\begin{eqnarray}
H_a =&& K^a_a - K^{a+1}_{a+1} , \quad a = 1, \cdots , 8 ,
\quad\quad H_{11} = K^9_9 - K^{10}_{10} \cr H_{9} =&& K^9_9 +
K^{10}_{10} + R_1 - \quart \sum_{a=1}^{10} K^a_a , \quad H_{10} =
- 2 R_1 \quad . \label{eq:cartib}
\end{eqnarray}
As in the case of the mIIA theory, excluding the ${\bf R}_s$
generator of $GL(10,{\bf R})$ and the translations, it can be
verified by inspection of the commutator algebra given in Eqs.\
(\ref{eq:extrab}) and (\ref{eq:strb}), that this choice of
Chevalley basis suffices to generate the remaining $p$-form
generators of ${\cal G}_{IIB}$ algebra. We can define the
nonlinear realization as before, identifying a corresponding group
element for each of the $(p+1)$-form generators in the IIB theory,
coupling to supergravity pbranes. As shown by West \cite{w4}, the
analysis permits a successful deduction of the line elements of
the full spectrum of type II branes: the p=1, 5 branes of the
NS-NS sector and p=-1,1,3,5,7 Dbranes of the R-R sector. The NS-NS
p=7 solution simply recovers Minkowskian spacetime. \vskip 0.4in

\noindent{\large{\bf Appendix C: Spacetime Reduction and Spacetime
Emergence}

\vskip 0.1in Notice that from the perspective of $E_{11}$ as
illustrated in West's analysis in \cite{w4}, there is nothing to
distinguish 11d supergravity from the ten-dimensional IIA and IIB
theories: all three share the same rank eleven symmetry algebra.
To see that the latter two are {\em ten-dimensional} field
theories requires that we introduce the notion of {\em spacetime}:
in the purely algebraic formulation of $E_8^{(3)}$ proposed by
West \cite{west2,w4}, there are no spacetime translation
generators, ${\cal P}_a$, to begin with.

\vskip 0.1in There are three opposing suggestions for how to
introduce spacetime into the formalism of nonlinear realizations
\cite{w4}. In the context of $E_{8}^{(2)}$, also a competing
proposal for the symmetry algebra of M theory \cite{ganor},
Damour, Henneaux, and Nicolai \cite{dhn} proposed that the fields
in the nonlinear realization should be taken as functions of time
alone. Spatial dependence would arise thru the action of certain
higher level generators of the $E_8^{(2)}$ algebra, which had the
correct commutation relations to be identified as spatial
derivatives. Thus, \cite{dhn} showed that the 3-form, 6-form, and
dual graviton representations appearing at level 1, 2, and 3, of
$E_8^{(2)}$ contain tensors with the index structure of a $k$th
spatial derivative at levels $1+3k$, $2+3k$, and $3+3k$. It is
expected that similar identifications can be made for $E_8^{(3)}$
\cite{kw}. An alternative viewpoint has been put forward by West
\cite{west1}, namely, that spacetime can be incorporated by
constructing the nonlinear realization of the semi-direct product
of $E_8^{(3)}$ with a particular lowest weight representation
denoted ${\bar{l}}_1$: the ${\bar{l}}_1$ representation would
contain the coordinates of spacetime, but it should be noted that
it also contains coordinates corresponding to the central charges
of the supergravity theory. Details can be found in \cite{kw}. A
third suggestion comes from Englert and Houart \cite{eh}, who
propose that the fields in the nonlinear realization depend upon
an auxiliary parameter, which is extended to a full spacetime by
the identification of generators corresponding to spatial
derivative operators in the very-extended algebra, analogous to
the proposal of \cite{dhn}.

\vskip 0.1in Since all of these proposals are at a preliminary
stage of investigation, it behooves us to keep an open mind. On
the other hand, it should be noted that our nonperturbative
proposal for M Theory \cite{mat1,mtheory} dovetails neatly with
the algebraic formalism of hidden symmetries: identify the hidden
symmetry algebra with the extended symmetry algebra of a reduced
unitary matrix model that implements local symmetries. The large
$N$ limit of such a unitary matrix model naturally provides for
the emergence of the coordinates of a spacetime continuum as
shown
in \cite{mtheory}. Let us review this important result.

\vskip 0.1in We will begin by explaining our modified prescription
for planar reduction such that one obtains non-trivial reduced
matrix models from the planar reduction of gravity theories.
Consider introducing a {\em flavor} quantum number in the
nonmaximal 10d Einstein-Yang-Mills Lagrangian
\cite{fs,brwn,br2,br,mtheory}, replacing the gravitational
spin-connection and Yang-Mills vector potential with
$N$$\times$$N$ matrix arrays as follows:
\begin{eqnarray}
{\cal R} \to&&  \partial_{a} (\omega_{b}^{ab})_{AB} -
\partial_{b} (\omega_{a}^{ab})_{AB} + (\omega_{b}^{ac})_{AC}
(\omega_{a c}^b)_{CB}  - (\omega_{a}^{ac})_{AC} (\omega_{b
c}^b)_{CB}  \cr F^i_{ab} \to&&
\partial_{a} (A^i_{b})_{AB} - \partial_{b} (A^i_{a})_{ AB} +
f^{ijk} (A^j_{a})_{ AC } (A^k_{b})_{ CB}  \quad ,
\label{eq:ymlarge}
\end{eqnarray}
where the indices run from $ i$$=$$1$, $\cdots$, ${\rm dim}~ G$,
and $A,B$$=$$1$, $\cdots$, $ N$. We will need to include a trace
over the $U(N)$ flavor group in order that the new Lagrangian
density transform as a $U(N)$ singlet. This Lagrangian will have
the symmetry group $U(N)$$\times$$G$, except that the $U(N)$ is
{\em not} a gauge symmetry. Rather, it plays the role of a flavor
group. How does one give meaning to the planar reduction of a
locally supersymmetric Lagrangian with a {\em huge} flavor
symmetry group to a single spacetime point? And why have we
introduced a large $N$ flavor symmetry, as opposed to the usual
large $N$ gauge symmetries invoked in \cite{ek,bfss,ikkt}?

\vskip 0.1in Notice that if we were to carry out the large $N$
extension in analogy with the planar reductions of rigid
Yang-Mills Lagrangians \cite{ek,ikkt}, namely, replace the
anomaly-free Yang-Mills group with the unitary large $N$ group:
$SO(32)$ $\to$ $U(N)$, where $SO(32)$ $\subset$ $U(32)$ $\subset$
$U(N)$, and where $U(N)$ is a fully {\em gauged} symmetry, we
would find nothing of interest in the gravity sector of the
Lagrangian. Suppressing all spacetime derivatives in the
supergravity-Yang-Mills Lagrangian as usual, we obtain the
standard quartic unitary matrix potential in the Yang-Mills
sector: $[A_{\mu},A_{\nu}]^2$, where $A$ is now a matrix of rank
$N$, but the Einstein sector yields an uninteresting finite rank
correction to these terms. Thus, since the zehnbein and spin
connection are {\em finite}-dimensional matrix arrays when reduced
to a single spacetime point, dimensional reduction of the Einstein
action gives a {\em finite} number of terms of the general form,
$E^{\mu}_a \omega^{ac}_{\mu} \omega^{b}_{c\lambda} E^{\lambda}_b
$. It is clear that if one desires a nontrivial modification of
large $N$ dynamics of the Yang-Mills sector {\em the gravity
variables must also scale with $N$}. This implies that the role of
$U(N)$ in the continuum field theory Lagrangian must be that of a
{\em flavor} symmetry group, rather than of a gauge group. We will
find that the introduction of $U(N)$ as a flavor symmetry in the
continuum Lagrangian enables the democratic appearance of large
$N$ scaling behavior in both the gauge and gravity sectors of the
matrix model obtained upon spacetime reduction to a single point.

\vskip 0.1in This motivates the first innovation introduced by us
in \cite{mat1,mtheory}: both the vector potential, zehnbein, and
spin connection, were required to transform in the adjoint
representation of a large $N$ flavor group. The same requirement
was made of the other bosonic fields in the Lagrangian, namely,
the dilaton and two-form potential. What about the spinors in the
supergravity Lagrangian? Since we have required that the large $N$
flavor group commute with the group of supersymmetry
transformations, it is important that fields which are partners
under supersymmetry belong to the same $U(N)$ representation.
Thus, we will require that all spinor fields gravitino, dilatino,
and gaugino, also transform in the adjoint representation of the
large $N$ flavor group. Keeping only the terms in the continuum
Lagrangian that remain after planar reduction, gives the following
supermatrix Lagrangian:\footnote{We denote the spacetime field,
$f(x)$, and its planar-reduced representative which lives at the
origin of spacetime, $f(0)$, by the same symbol $f$.}
\begin{eqnarray}
{\cal L}^{(10d)}_{\rm planar} &&=~ \half \gap {\rm tr} \left (
f^{ijk} f^{ilm} A^{l a} A^{m b} A^j_{a} A^k_{b} ~+~ {\bar \chi}^i
\Gamma^a A_{a}^{j} \tau^{j} \chi^i \right ) \cr \quad\quad &&\quad
~+ \kap {\rm tr} \left ( E_a^{\mu} \left [ \omega_{\nu}^{ac}
\omega_{\mu c}^b  - \omega_{\mu}^{ac} \omega_{\nu c}^b \right ]
E_{b}^{\nu} - A^i_{a} \tau^i \Phi A_{a}^j \tau^j \Phi  + {\bar
\chi}^i \Gamma^a \omega_{abc}\Gamma^{bc} \chi^i \right ) \cr \quad
&&\quad\quad + \kap {\rm tr} \left ( {\bar \psi}_{a} \Gamma^{abc}
\omega_{bde}\Gamma^{de} \psi_{c} - 4 {\bar{\lambda}} \Gamma^{ab}
\omega_{ade}\Gamma^{de} \psi_{b} - 4 {\bar{\lambda}} \Gamma^{a}
\omega_{ade}\Gamma^{de} \lambda \right ) \cr \quad &&
\quad\quad\quad + {{4}\over{3}} {{1}\over{g^4}} {\rm tr} \left (
f^{nlm} f^{ijk} A^{n[a} A^{l b} A^{m c ]} A^i_{[a} A^j_{b} A^k_{c
]} \right )
   + {\rm tr} ~{\cal L}_{\rm 2-fermi} ~+~ {\rm tr} ~{\cal L}_{\rm
4-fermi}  \quad . \label{eq:lmat}
\end{eqnarray}
where $i,j,k, \cdots $ are group indices for the
finite-dimensional Yang-Mills gauge group, and repeated indices
are to be summed. The notation \lq\lq tr" denotes, instead, the
trace over the large $N$ flavor group, whose indices have been
suppressed. The first line of this expression is familiar:
analogous terms appear in both the Banks-Fischler-Shenker-Susskind
\cite{bfss} and Ishibashi-Kawai-Kitazawa-Tsuchiya \cite{ikkt}
rigid matrix models. The index structure of the terms in the first
line make the $U(N)$$\times$$G$ symmetry of the supermatrix model
manifest: the model obtained by restricting to only the terms in
the first line of this expression defines the {\em simplest}
possible supermatrix model consistent with this symmetry group.

\vskip 0.1in Thus, the distinction between flavor, and gauged,
large $N$ symmetry becomes significant only when we take into
account the remaining terms in the matrix Lagrangian: we find new
large $N$ matrix variables originating in the supergravity sector
of the continuum Lagrangian, as well as new multi-matrix
interaction terms. These include a sixth-order self-interaction
for the Yang-Mills potential, a term which was absent in both the
BFSS and IKKT matrix models \cite{bfss,ikkt}, and which arises
from the Chern-Simons contribution to the supergravity three-form
field strength. It is evident that the symmetry structure of the
full supermatrix Lagrangian given in Eq.\ (\ref{eq:lmat}) is much
more subtle than simply $U(N)$$\times$$G$. Knowledge of the
Cremmer-Julia hidden symmetries of the continuum field theory
Lagrangian becomes a useful tool for its analysis.

\vskip 0.1in It is illuminating to examine the form of matrix
Lagrangian obtained by the planar reduction of the 11d
supergravity theory. Recall the absence of Yang-Mills gauge
fields, as well as the absence of a dilaton supermultiplet, in 11d
supergravity. The 11d supergravity theory does, however, include a
four-form field strength. The associated three-form potential
couples to the supermembrane. Introduction of a large $N$ flavor
quantum number in the continuum Lagrangian, followed by spacetime
reduction of the field theory to a single spacetime point, gives
an elegant and especially simple matrix model:
\begin{eqnarray}
{\cal L}^{(11d)}_{\rm planar} =&& \kap {\rm tr} \left \{ {\bar
\psi}_{a} \Gamma^{abc} \omega_{bde} \Gamma^{de} \psi_{c} +
\omega_{b}^{ac} \omega_{a c}^b - \omega_{a}^{ac} \omega_{b c}^b
\right \} \quad , \label{eq:11dmat}
\end{eqnarray}
where the gravitino, $\psi_a$, is a 32-component Grassmann-valued
array, and $\omega_{abc}$ is the spin-connection. Notice that the
the presence of a higher p-form supergravity potential in the
continuum field theory Lagrangian is, unfortunately, erased from
the planar reduced matrix model: there is no analogous
Chern-Simons coupling to a Yang-Mills field, as was present in
Eq.\ (\ref{eq:lmat}). It may well be true that this particular
supermatrix Lagrangian falls within the class of solvable
zero-dimensional multi-matrix models, enabling a detailed analysis
by well-established matrix model techniques. We should emphasize
that this model is the precise, pure gravitational, supermatrix
model analog of the planar reductions of rigid supersymmetric
Yang-Mills theory considered in \cite{ek,bfss,ikkt}. However, the
model appears not to capture the full content of M theory because
it lacks any knowledge of the crucial supermembrane sector of the
theory.

\vskip 0.1in Thus, while the planar reduction of gravity theories
with large $N$ flavor group has led to an interesting new class of
zero-dimensional matrix models, these models appear not to capture
the full content of M theory, inclusive of the crucial
brane-spectrum required by duality. This brings us to a second
innovation introduced in \cite{mat1}. Notice that the Lagrangian
considered in \cite{br2,br,mat1,mtheory} describes a
ten-dimensional supergravity theory in generic curved spacetime
background. Inherent in this expression is the notion of a local
ten-dimensional flat tangent space attached to every point in
spacetime. The naive procedure of planar reduction we have
borrowed from rigid super-Yang-Mills theories \cite{ek} has
ignored this aspect of the supergravity Lagrangian. We will now
show that the spacetime reduction of all spacetime fields to {\em
linear} forms defined on the infinitesimal patch of local tangent
space at a single spacetime point, suffices to ensure that all of
the local symmetries of the continuum Lagrangian are preserved in
a corresponding reduced matrix model.

\vskip 0.1in Our starting point is the unusual field theory
Lagrangian with a huge flavor symmetry group: all fields, bosonic
and fermionic, are required, in addition, to live in the adjoint
representation of the large $N$ unitary group, $U(N)$. We
emphasize that $U(N)$ is a flavor symmetry; only the finite rank
anomaly-free Yang-Mills group $G$ has been gauged. Starting with
the nonmaximal d=10 supergravity theory coupled to $O(32)$
Yang-Mills fields, we have the corresponding $U(N)$ invariant
Lagrangian:
\begin{eqnarray}
{\cal L} &&=~ \kap {\rm tr} \left \{ {\bar \psi}_{a} \Gamma^{abc}
D_{b} (\omega) \psi_{c} ~-~ 4 {\bar{\lambda}} \Gamma^{ab} D_{a}
(\omega) \psi_{b} ~-~ 4 {\bar{\lambda}} \Gamma^{a} D_{a} (\omega)
\lambda \right \} \cr \quad &&\quad +~ \gap ~ {\rm tr} \left \{
{\bar \chi}^i \Gamma^{a} D_{a} (\omega,A) \chi^i~+~ \half ~
{F}^{ab}(A) F_{ab} (A) \right \} \cr \quad\quad &&\quad\quad ~+~
\kap {\rm tr} \left \{ {\cal R}(\omega,E) ~-~
\partial^{a} \Phi ~\partial_{a} \Phi ~+~ 3 {H}^{abc} {H}_{abc}
\right \} \cr && \quad\quad\quad\quad~+~ {\rm tr} \left \{ {\cal
L}_{\rm 2-fermi} ~+~ {\cal L}_{\rm 4-fermi}  \right \} \quad .
\label{eq:hmatflav}
\end{eqnarray}
where the notation \lq\lq tr" denotes taking the trace over the
large $N$ flavor group, and the two- and four-fermi terms are as
given in \cite{br,mat1,mtheory}. Notice that each term in the
Lagrangian is a flavor singlet, and the $U(N)$ flavor group
commutes with {\em all} of the spacetime symmetries of the
Lagrangian: namely, local Lorentz and local supersymmetry
transformations, in addition to Yang-Mills gauge
transformations.\footnote{In our earlier papers \cite{mat1}, we
have pointed out a more general possibility for the matrix
superalgebra. Namely, the parameters for infinitesimal
supersymmetry and $SL(n,{\bf R})$ transformations could themselves
be non-singlet under the flavor $U(N)$. While we know of no reason
to rule out such an extension, it is not necessary for the problem
at hand. Notice that, for such matrix algebras, the large $N$
limit would have to correspond to an exotic (nonlinear) extension
of the Nahm classification of spacetime linear superalgebras. We
thank Bernard de Witt for pointing this out.}

\vskip 0.1in The $E_a^{\mu}$ are the fundamental variables
appearing in the matrix Lagrangian, but they are not all
independent. Assuming a flat tangent space of Minkowskian
signature, $\eta_{ab}$, the usual relation for the spacetime
metric tensor takes the form of a $U(N)$ identity:
\begin{equation}
G^{\mu\nu} = {\rm tr} \left ( E^{\mu}_a E^{\nu}_b \right ) , \quad
\mu,\nu, ~ {\rm and} ~ a,b=0, \cdots 9 , \quad\quad E^{\mu}_a =
G^{\mu\nu} E_{\nu a} \quad . \label{eq:basicsc}
\end{equation}
As is familiar from differential geometry, $G^{\mu\nu}$ is the
object that raises spacetime indices, while $\eta^{ab}$ is the
object that raises indices in tangent space. The spacetime metric
transforms as a $U(N)$ singlet, as does $\eta^{ab}$. The usual
constraint equation relating them is automatically satisfied:
\begin{equation}
\eta_{ab} = {\rm tr} \left ( E^{\mu}_{ a} E_{\mu b} \right ) =
G^{\mu\nu} {\rm tr} \left ( E_{\nu a } E_{\mu b} \right ) =
G^{\mu\nu} {\rm tr} \left ( G_{\nu \lambda} \eta_{ac} E^{\lambda
c} E_{\mu b} \right ) = \delta^{\mu}_{\lambda} {\rm tr} \left (
\eta_{ac} E^{\lambda c} E_{\mu b} \right ) = \eta_{ab} \quad .
\label{eq:constrmet}
\end{equation}

\vskip 0.1in We will now reduce all spacetime fields to {\em
linear} forms defined on the infinitesimal patch of local tangent
space at a single spacetime point as explained above. In other
words, instead of simply setting all spacetime derivatives to zero
as in the previous section, we retain the $O(\delta \xi^a)$ terms
of the continuum Lagrangian, truncating at $O((\delta \xi^a)^2)$
in the Taylor expansion on tangent space, in the infinitesimal
vicinity of the spacetime origin. We have parameterized the
infinitesimal patch of local tangent space at the origin by the
variables $\xi^a $, $a$$=$$0$, $\cdots$, $9$. Recall the usual
relation in Riemannian differential geometry linking the partial
derivative operators acting in spacetime, and in the local tangent
space:
\begin{equation}
\partial_{\mu} = E_{\mu}^a \partial_a , \quad \quad \mu , \nu = 0, \cdots , 9,
\quad a,b = 0, \cdots , 9 \quad , \label{eq:basics}
\end{equation}
Since the zehnbein is a flavor adjoint, an $N$$\times$$N$
dimensional array, whereas $\partial/\partial \xi^a$ is the
ordinary partial derivative operator acting on a continuous and
differentiable space with the local geometry of $R^{10}$, {\em it
follows that the partial derivative operator in spacetime,
$\partial_{\mu}$, is also $U(N)$ valued}. In particular,
consistency with the obvious identity $\partial_{\mu}
X^{\mu}$$=$$1$, implies that:
\begin{equation}
( X^{\mu})_{AB} \equiv (E^{\mu}_{ a})_{AB} \delta \xi^a , \quad
\quad (\partial_{\mu})_{AB} (X^{\mu})_{BC} = ({\bf 1})_{AC} ,
\quad \quad A,B =1, \cdots N \quad . \label{eq:vielb}
\end{equation}
In other words, the coordinate vector, $X^{\mu}$, is itself $U(N)$
valued! Notice that $X^{\mu}$ is a dependent variable in our
framework: it is {\em derived} from the zehnbein, $E_{\mu}^a$,
which is the fundamental variable appearing in the matrix
Lagrangian. Unlike tangent space, which is smooth and
differentiable, at least infinitesimally, spacetime contains a
single element, a single spacetime \lq\lq point". All variables
defined at this single spacetime point are $N$$\times$$N$
dimensional matrices, the fundamental degrees of freedom in the
matrix model Lagrangian. Further, we will require of all forms on
tangent space that they satisfy the {\em linearity} property:
quadratic, and higher, derivatives, are identically set to zero,
$\partial_n f(0)$ $=$ $0$, $n\ge2$, reflecting the fact that we
are defining a set of functions on a base manifold of {\em
infinitesimal} extent.

\vskip 0.1in Why is there a need to retain an infinitesimal patch
of tangent space while performing the dimensional reduction of the
gravity theory to a single spacetime point? To develop some
intuition into the $U(N)$ valued relations given above, notice
that no restrictions have been placed upon the eigenvalue spectrum
of the various zehnbein. In principle, one can solve for the
eigenvalue spectrum of each $E^{\mu}_a$, given the equation of
motion that follows from the classical matrix Lagrangian. One of
the solutions to the equation of motion corresponds to choosing
the 10d Minkowskian flat space time metric as classical
background:
\begin{equation}
<G^{\mu\nu}> ~=~ <{\rm tr} \left ( E^{\mu}_a E^{\nu}_b \right )>
~=~ \eta^{\mu\nu} , \quad \mu,\nu, ~ {\rm and} ~ a,b=0, \cdots 9
\quad . \label{eq:basicflat}
\end{equation}
We solve for the corresponding $<E_{\mu}^a>$, expressing them in
diagonal form, and ordering the eigenvalues along the diagonal to
reflect a {\em monotonic increase}. It is evident that in the
large $N$ limit, the eigenvalues will crowd together forming a
continuum. Of course, as a consequence of the identity in Eq.\
(\ref{eq:vielb}), the coordinate matrices, $<X^{\mu}>$, also take
diagonal form, their entries reflecting the monotonic increase
along the diagonals of individual zehnbein. It is natural to
interpret the ordered continuum of eigenvalues of the coordinate
matrix as coordinate-locations for the continuum of spacetime
points along the coordinate axis $x^{\mu}$ of 10d Minkowskian
spacetime. Thus, we have recovered the coordinates of the
spacetime continuum by taking the large $N$ limit of the matrix
model!

\vskip 0.1in  We are now ready to carry out the spacetime
reduction of the Lagrangian given in Eq.\ (\ref{eq:hmatflav}) in
accordance with our new prescription. Note that all fields,
bosonic or fermionic, transform as adjoints under the flavor
$U(N)$, and every term in the Lagrangian is a $U(N)$ singlet. The
Lagrangian is manifestly invariant under local supersymmetry and
local Lorentz transformations, and these symmetries commute with
the flavor $U(N)$. Under spacetime reduction, every field in the
Lagrangian is reduced to at most a {\em linear} form on the local
tangent space, reflecting the fact that tangent space is an {\em
infinitesimal} manifold. The sole exception is the zehnbein: since
the spacetime coordinates have turned out to be in one-to-one
correspondence with the eigenvalue spectrum of the zehnbein in the
large $N$ limit, self-consistency requires that the reduction of
the zehnbein is to a {\em zero-form} on tangent space. Most
importantly, this also has the natural consequence that the local
symmetries of the continuum Lagrangian can be made manifest in the
reduced matrix model.

\vskip 0.1in The remaining independent dynamical fields in the
Lagrangian reduce to {\em linear} forms on tangent space. Our
notation for a generic one-form $f(\xi)$ is as follows: $f(\xi)$
$=$ $f(0)$ $+$ $\partial_a f(0)$ $ \delta \xi^{a}$, where
$\partial_a f(0)$ denotes, more precisely, the partial derivative
of $f$ with respect to $\xi^a$, evaluated at $\xi^a$ $=$ $0$.
Since every field in the continuum Lagrangian is an $N$$\times$$N$
array under flavor $U(N)$, and global symmetries are preserved
under spacetime reduction, $f(0)$ and $\partial_a f(0)$ are two
{\em independent} unitary matrices appearing in the matrix
Lagrangian. Of course, one or other matrix array will be found to
drop out of any given term in the Lagrangian. Thus, the matrix
Lagrangian will turn out to have {\em exactly} the same symmetry
group as the original continuum field theory Lagrangian. Listing
each of the independent dynamical fields appearing in Eq.\
(\ref{eq:hmatflav}), we have the following result upon spacetime
reduction to corresponding $N$$\times$$N$ matrix arrays defined at
the origin $x$ $=$ $0$, which we choose coincident with the origin
of tangent space, $\xi$ $=$ $0$:
\begin{eqnarray}
E^{\mu}_a (x) &&\to E^{\mu}_a (0)  \cr A_a^i (x) &&\to A_a^i(0) +
\partial_b A_{a}^i(0) \delta \xi^b  \cr
 \partial_{c} A_{d}^i (x)  &&\to \partial_c (A_d^i(0)) + \partial_b A_{d}^i (0) \partial_c ( \delta \xi^b )
  = \partial_c A_{d}^i (0) \cr
D_c A_d^i (x) &&\to \partial_c A_{d}^i(0)  - \partial_d A_{c}^i
(0) + f^{ijk} A_c^j(0) A_d^k(0) \cr
 H_{abc} (x) &&\to
\partial_{[c} B_{ab]} (0) - g^2\left ( A^i_{[a} (0)
\partial_{b} A^i_{c]} (0)
  - \twothird f^{ijk} A^i_{[a}(0) A^j_{b}(0) A^k_{c ]}(0) \right ) \cr
  D_a \Phi (x) && \to \partial_a \Phi (0) - A_a^i \tau^i \Phi(0)
\quad , \label{eq:formsf}
\end{eqnarray}
where the indices have range as follows: $\mu $ $=$ $0$, $\cdots$,
$9$, $a,b,c$ $=$ $0$, $\cdots$, $9$, and $ i,j,k$ $=$ $1$,
$\cdots$, ${\rm dim} ~ G$. We remind the reader that each of the
objects on the left-hand-side of this list is also an
$N$$\times$$N$ unitary matrix, the flavor indices have simply been
suppressed. Suppressing the \lq\lq$(0)$" dependence, we obtain the
matrix Lagrangian:
\begin{eqnarray}
{\cal L}^{\rm (mat)} &&=~ \kap {\rm tr} \left \{ {\bar \psi}_{a}
\Gamma^{abc} D_{b} (\omega) \psi_{c} ~-~ 4 {\bar{\lambda}}
\Gamma^{ab} D_{a} (\omega) \psi_{b} ~-~ 4 {\bar{\lambda}}
\Gamma^{a} D_{a} (\omega) \lambda \right \} \cr && \quad ~+~ \kap
{\rm tr} \left \{ {\cal R}(\omega,E) ~-~ \partial^{a} \Phi
~\partial_{a} \Phi ~+~ 3 {H}^{abc} {H}_{abc} \right \}
\cr&&\quad\quad ~+~ \gap {\rm tr} \left \{ \half ~ {F}^{ab}(A)
F_{ab} (A) + ~ {\bar \chi}^i \Gamma^{a} D_{a} (\omega,A) \chi^i
\right \} \cr \quad && \quad\quad\quad\quad ~+~ {\cal L}^{\rm
(mat)}_{\rm 2-fermi} ~+~ {\cal L}^{\rm (mat)}_{\rm 4-fermi}  \quad
. \label{eq:hmatfm}
\end{eqnarray}
In other words, the matrix Lagrangian takes precisely the {\em
same} form as the original continuum Lagrangian with large $N$
flavor group, except that all spacetime fields are restricted to
their value at the origin: the infinite number of degrees of
freedom in the original continuum field theory have indeed been
drastically thinned to those of a zero-dimensional matrix model
with $U(N)$ flavor symmetry. But, remarkably, by the introduction
of infinitesimal linear forms on the local flat tangent space,
this matrix Lagrangian also preserves a remnant of the {\em local}
symmetries of the continuum Lagrangian. The underlying reason why
there exists a matrix Lagrangian that can make manifest the local
symmetries of a given continuum field theory, is the spacetime
{\em locality} property of the Lagrangian density in quantum field
theory. Further details, including the crucial two- and four-fermi
terms required by closure of the supersymmetry algebra can be
found in \cite{mat1,mtheory}, following \cite{br2,br}.

\end{document}